\documentclass[conference,compsoc]{IEEEtran}
\usepackage{fancyhdr}
\ifCLASSOPTIONcompsoc
  \usepackage[nocompress]{cite}
\else
  \usepackage{cite}
\fi
\ifCLASSINFOpdf
\else
\fi
\usepackage[normalem]{ulem}
\usepackage[english]{babel}
\usepackage{amsthm}

\usepackage{color}
\usepackage[utf8]{inputenc}
\usepackage{fancyvrb}
\usepackage{xcolor}
\usepackage{graphicx}
\usepackage{comment}

\usepackage{caption}
\usepackage{subcaption}

\usepackage{booktabs} 
\usepackage{url}

\usepackage{algorithm} 
\usepackage{algpseudocode} 
\usepackage{mathtools}

\usepackage{multirow}
\theoremstyle{definition}

\newcommand{\etal}{\textit{et al}.}

\newcommand{\fixit}[1] {
  {\color{red}{#1}}}

\begin{document}

%
\title{Optimizing Quantum Fourier Transformation (QFT)  Kernels  \\ for Modern NISQ and FT Architectures}

\author{

\IEEEauthorblockN{ Yuwei Jin*}
\IEEEauthorblockA{
\textit{Rutgers University}\\
Piscataway, USA\\
jyw413482880@gmail.com}

\and

\IEEEauthorblockN{Xiangyu Gao*}
\IEEEauthorblockA{
\textit{New York University}\\
New York, USA\\
xg673@nyu.edu}
\and
\IEEEauthorblockN{ Minghao Guo}
\IEEEauthorblockA{\textit{Rutgers University}\\
Piscataway, USA\\
mg1998@cs.rutgers.edu}
\and
\IEEEauthorblockN{ Henry Chen}
\IEEEauthorblockA{\textit{Rutgers University}\\
Piscataway, USA\\
hc867@scarletmail.rutgers.edu}
\and
\IEEEauthorblockN{ \hspace{2cm}Fei Hua}
\IEEEauthorblockA{\hspace{2cm}\textit{Rutgers University}\\
\hspace{2cm}Piscataway, USA\\
\hspace{2cm}huafei90@gmail.com}
\and
\IEEEauthorblockN{ Chi Zhang}
\IEEEauthorblockA{\textit{Independent Researcher}\\
USA\\
raymond.chizhang@gmail.com
}
\and
\IEEEauthorblockN{ Eddy Z. Zhang}
\IEEEauthorblockA{\textit{Rutgers University}\\
Piscataway, USA\\
eddy.zhengzhang@gmail.com}

}

\maketitle



\begin{abstract}

Rapid development in quantum computing leads to the appearance of several quantum applications. Quantum Fourier Transformation (QFT) sits at the heart of many of these applications. 
Existing work leverages SAT solver or heuristics to generate a hardware-compliant circuit for QFT by inserting SWAP gates to remap logical qubits to physical qubits. However, they might face problems such as long compilation time due to the huge search space for SAT solver or suboptimal outcome in terms of the number of cycles to finish all gate operations. In this paper, we propose a domain-specific hardware mapping approach for QFT. 
We unify our insight of relaxed ordering and unit exploration in QFT to search for a qubit mapping solution with the help of program synthesis tools. Our method is the first one that guarantees linear-depth QFT circuits for Google Sycamore, IBM heavy-hex, and the lattice surgery, with respect to the number of qubits.
Compared with state-of-the-art approaches, our method can save up to $53\%$ in SWAP gate and $92\%$ in depth. \footnote{ * These authors contributed equally to this work.}


\end{abstract}


%
\IEEEpeerreviewmaketitle

\section{Introduction}

Quantum computing has exhibited its pronounced advantages in several fields, including cryptography \cite{shor:focs94}, financial modeling \cite{rebentrost+:PhysRevA18}, and chemical simulations \cite{peruzzo+:ncomms14}.

Quantum Fourier Transform (QFT) is a computation kernel used in many important quantum applications \cite{coppersmith:qft} \cite{bova2021commercial}.
The applications that utilize the QFT kernel include but are not limited to the following: Shor's algorithm \cite{shor:focs94}, amplitude estimation algorithm \cite{rebentrost+:PhysRevA18} \cite{walter+:monteCarlo},  quantum phase estimation (QPE) \cite{kitaev:QPE},  hidden subgroup problem (HSP) \cite{ettinger1999:hsp}, HHL algorithm \cite{harrow_2009:hhl}, and quantum walks \cite{childs_2003:quantumwalk}. These algorithms are fundamental for many domains, including cryptography, finance, risk analysis, quantum simulations, and chemistry and materials sciences. These algorithms include both \emph{near-term} and \emph{long-term} applications. They cover not only long-term applications such as Shor's algorithm, which requires a significant number of qubits and gates, but also near-term applications such as QPE, which requires a moderate number of qubits and gate resources.  

\begin{figure}[htb]
    \centering
    \includegraphics[width=0.4\textwidth]{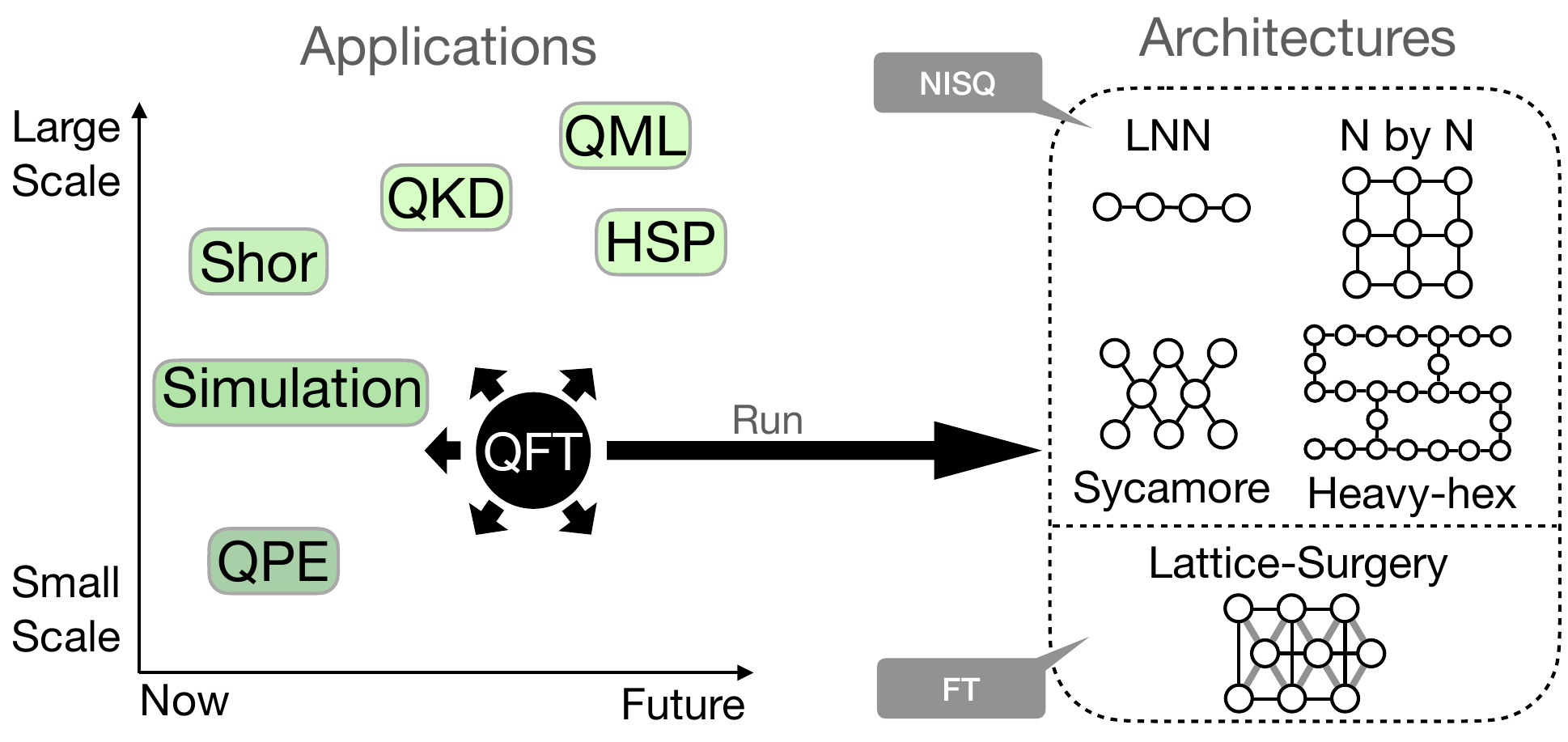}
    \caption{QFT serves as a computation kernel of diverse applications (e.g., Quantum Phase Estimation(QPE), Quantum Key Distribution(QKD), Quantum Machine Learning(QML), and Hidden Subgroup Problem(HSP)). There is a need to optimize QFT over diverse backends on the RHS.}
    \label{fig:diversebackends}
\end{figure}

Hence, optimizing the QFT kernel is of utmost importance. However, it is challenging for the following reasons. 

\textbf{{(1)}} The QFT kernel is complex due to its requirement for all-to-all qubit interactions and strict dependency constraints, as the example shown in Fig.~\ref{fig:qftlogical}. Each qubit must interact with every other qubit using a controlled phase rotation (CPHASE) gate. Each qubit must act either as the control qubit or target qubit in the CPHASE gate, and follow a strict of ordering being control and target qubits.  This complexity increases as the qubit number increases.

\textbf{{(2)}} The QFT kernel and other well-known quantum algorithms are often designed or optimized without specific hardware constraints in mind. Hence it is important to efficiently compile the logical QFT kernels into hardware circuits. For instance, the sparsity in today's hardware connectivity significantly hampers the execution of long-distance two-qubit gates. Two qubits must be physically adjacent to each other to execute a 2-qubit gate. This issue exists for both the noisy intermediate-scale quantum (NISQ) backends and certain fault-tolerant (FT) backends.

\textbf{{(3)}} The real-world quantum computing environment has the issues of noises, scalability, limited connectivity, and diversity in the backends. A QFT kernel may need to run in heterogeneous backends. These backends include the  NISQ backends, such as the IBM heavy-hex device and the Google Sycamore device, and FT backends, such as the surface-code QEC devices equipped with the lattice surgery mode for 2-qubit gate implementations. It is challenging to provide efficient compilation solutions for a variety of backends and ensure consistent compilation quality. Moreover, the current compiler must recompile each time, given a different input size of the QFT kernel, and may fail to provide consistent quality across different input sizes.

\textbf{{Our Goal.}} Our paper focuses on optimizing the QFT kernel in the real quantum computing environment. As mentioned above, it is important to optimize QFT for heterogeneous backends, and have them ready before plugging them in different important quantum applications. We tackle this problem by leveraging the following key insights: 

\textbf{Key Insight 1: Break the Strict Ordering of Gates.}  The QFT kernel has abundant CPHASE gates. Each CPHASE gate commutes with the other. Hence, these gates do not have to follow the strict dependence order in the conventional logical circuits for the QFT kernel represented in the quantum textbooks \cite{qcqi_book}. Based on our experiment results, exploiting the commuting property does not affect the correctness of QFT, but makes a big difference when performing SWAP insertion during the compilation. 

\textbf{Key Insight 2: QFT Subkernel Partitioning.} We develop a novel and flexible QFT sub-kernel partition methodology. This methodology allows reducing the compilation problem of the QFT kernel for a high-dimensional architecture into that for a low-dimensional architecture.  

\textbf{Key Insight 3: Unifying Different Methods.} We unify the approaches from Insight 1 and Insight 2. Combining these two approaches can significantly improve the overall QFT kernel performance on hardware. For instance, while being able to reduce high-dimensional problem to low-dimensional problem, our sub-kernel splitting may cause extra delay in the circuit. However, breaking the ordering of gates may compensate for the delay of the circuits. We present a framework of methodology, and showcase multiple scenarios these two key insights can be flexibly combined.

\textbf{Other Insights.} We leverage program synthesis \cite{sketch} methods to help enhance the compilation of the QFT kernels. It helps integrate human intelligence (educated guesses) for finding structured solutions to exploit the \emph{regularity in both the QFT kernel and the scalable hardware we deal with}. Our experience in using program synthesis for compiling the QFT kernel is not only useful for QFT but may potentially be useful for compiling other quantum algorithms with a regular structure.

\textbf{Our Contributions.} With all the above, we have developed {a unified framework of methodology for compiling the QFT kernel}. Our framework has the following benefits: \emph{(1) Our approach does not require recompilation when the number of qubits changes}; \emph{(2) Our approach adapts to different backends, including both NISQ and FT backends;} \emph{ (3) Our approach provides consistent performance and fidelity guarantee for different input sizes and backends}.

Most importantly, for the first time, our work discovered \emph{linear-depth hardware QFT kernel} for NISQ devices including IBM heavy-hex and Google Sycamore, and FT devices equipped with the surface code QEC and the lattice surgery model.   Our experiments have demonstrated a huge improvement in both depth and gate count over the state-of-the-art approaches. Our contributions are multifold: 

\begin{itemize}
    \item Guaranteed compilation quality: linear-depth solutions for the QFT kernel on different architectures:  Google Sycamore, IBM Heavy-hex, and an FT backend with lattice surgery model.
    \item Commutativity exploitation for reordering CPHASE gates in the QFT kernel. 
    \item A novel and flexible QFT sub-kernel partition that allows hierarchical decomposition of high-dimensional problems to low-dimensional problems. 
\item In-depth understanding of the optimization space for the QFT kernel on real hardware. 
    \item Our hardware QFT kernel mapping solutions have up to $53\%$ fewer SWAP gate count and $92\%$ less depth than state-of-the-art approaches for up to 1024 qubits compared with SABRE \cite{li+:asplos19}.
\end{itemize}

\section{Background}
\label{sec:motivation}

We introduce the hardware mapping problem first. Then we introduce a low-dimensional QFT kernel. We also describe both the NISQ backends and FT backends, in Section \ref{sec:lowdim} and Section \ref{sec:ft}. respectively. 

{\subsection{Hardware Mapping}}
In superconducting quantum hardware, a two-qubit gate can only be executed when its two qubits are located in 2 adjacent physical qubits. In reality, due to the sparse connectivity of the current quantum machine, SWAP gates are required to change the hardware mapping and enable two-qubit gate interaction. For instance, for a line topology $q_0 \xleftrightarrow{} q_1 \xleftrightarrow{} q_2 \xleftrightarrow{} q_3 $, to execute a two-qubit gate between $q_0$ and $q_3$, one can insert 2 SWAP gates: SWAP($q_0$, $q_1$) and SWAP($q_2$, $q_3$)  to move q0 and q3 closer.

\subsection{A Low-dimensional QFT Kernel}
\label{sec:lowdim}

\begin{figure}[htb]
    \centering
\includegraphics[width=0.35\textwidth]{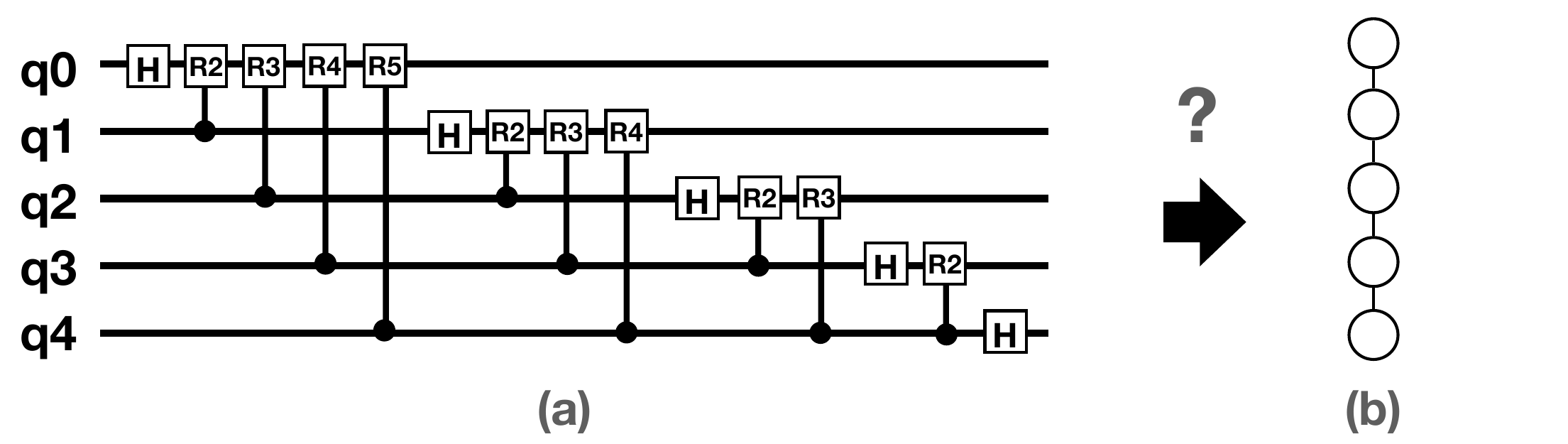}
    \caption{QFT Logical Circuit and a Line Topology}
    \label{fig:qftlogical}
\end{figure}

Prior studies \cite{maslov:physreva16} \cite{zhang+:asplos21} have discovered a hardware mapping solution for the QFT kernel on the linear nearest neighbor (LNN) architecture, which is a line of connected qubits. In our sub-kernel partition approach, we will recursively decompose the problem into the low-dimensional case, this LNN case serves as \emph{our base case for low-dimensional problem}. This base case also serves as an inspiration for our use of the program synthesis tool when discovering QFT kernel solution on non-trival architectures. The logical circuit of the QFT kernel is shown in Fig. \ref{fig:qftlogical} (a).

\begin{figure}[htb]
    \centering
\includegraphics[width=0.45\textwidth]{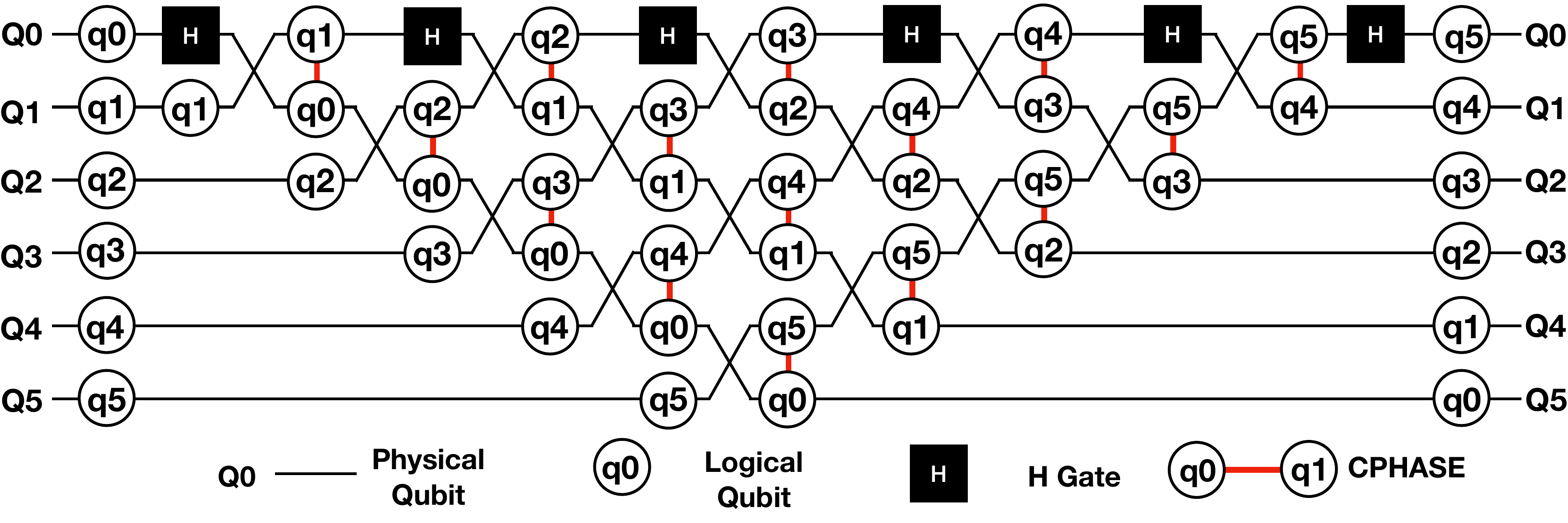}  
    \caption{Hardware mapped QFT in LNN. $q_i$ is a logical qubit. $Q_i$ is a physical qubit. The single-qubit gate runs in parallel with two-qubit gates. Each qubit moves to the top first and then moves down, except q0, which directly moves down. When a qubit is at the top, it stops for one time step. }      
    \label{fig:LNN5}
    \vspace{-5pt}
\end{figure}

We show the LNN solution via an example in Fig. \ref{fig:LNN5}.  The circuit clearly exhibits a pattern. We will analyze this pattern. Before the analysis, we assume N is the number of qubits, $q_i$ represents a logical qubit, and $Q_i$ represents a physical qubit.  

\textbf{Initial mapping:} Each logical qubit $q_i$ is mapped to the physical qubit $Q_i$ initially, $q_i \rightarrow Q_i$.

\textbf{Repeating Steps:} We denote a CPHASE gate between physical qubits $Q_i$ and $Q_{i+1}$ as $G(Q_i, Q_j)$, where $Q_j$ is the control in the CPHASE gate, and $Q_i$ is the target in the CPHASE gate. Each parallel layer is a sequence of consecutive pairs of SWAP or CPHASE starting from the physical qubit $Q_0$ or $Q_1$. The upper bound of the qubit index increases by one at one time for the physical qubits.  Note that, interestingly, for the i-th parallel layer of CPHASE or SWAP gates, the logical qubits in each pair, have their indices sum to a constant number for each layer.  At the very end, the qubits are as if reversed on the line, such that the mapping is $q_i$ $\rightarrow$ $Q_{N-1-i}$.  

\begin{figure}[htb]
    \centering
\includegraphics[width=0.28\textwidth]{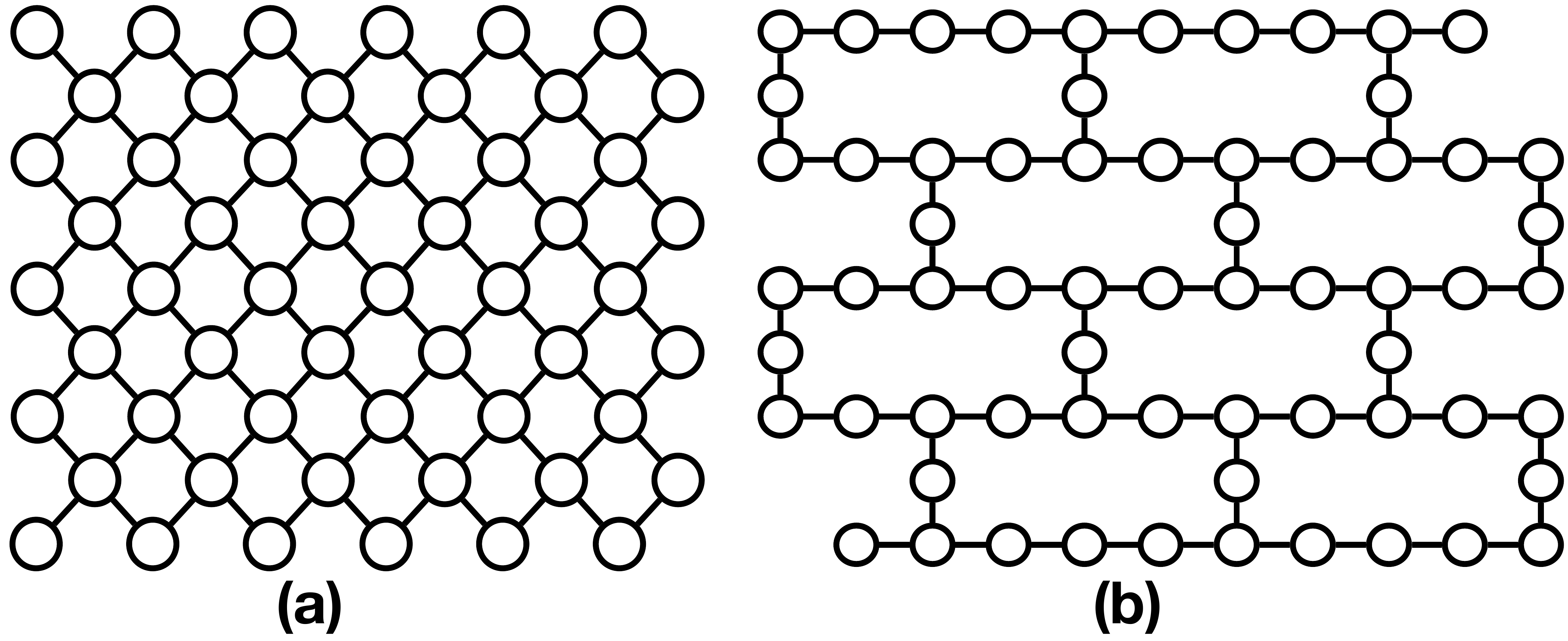}
    \caption{(a) Google Sycamore (b) Heavy-hex}
    \label{fig:modernarch}
\end{figure}

\textbf{Implications:} One may hope to find a Hamiltonian path that connects all the qubits and then apply the LNN solution for the QFT kernel. However, it is difficult to find such a path in modern architectures, whether NISQ or FT backends.  Fig. \ref{fig:modernarch} (a) and (b) show, respectively, Google's and IBM's superconducting architecture. It is possible to find a path, but such a path may not be able to connect all nodes. It can be proven that the Hamiltonian path does not exist. 
Furthermore, checking the existence of a Hamiltonian path in one graph is an NP-complete problem~\cite{Hamiltonian_path}, meaning the cost spent in finding one Hamiltonian path might exceed the gain to leverage previously proposed efficient qubit mapping over LNN architecture. 

Nonetheless, this LNN solution is still useful. Our proposed solutions, as described later, are either \textbf{non-trivial} extensions of the LNN solution or will use the LNN case as the base case when we perform recursive sub-kernel partitioning. 

\subsection{The Fault-tolerant Backend and the Lattice Surgery Model}
\label{sec:ft}

\begin{figure}[htb]
    \centering
    \includegraphics[width=0.38\textwidth]{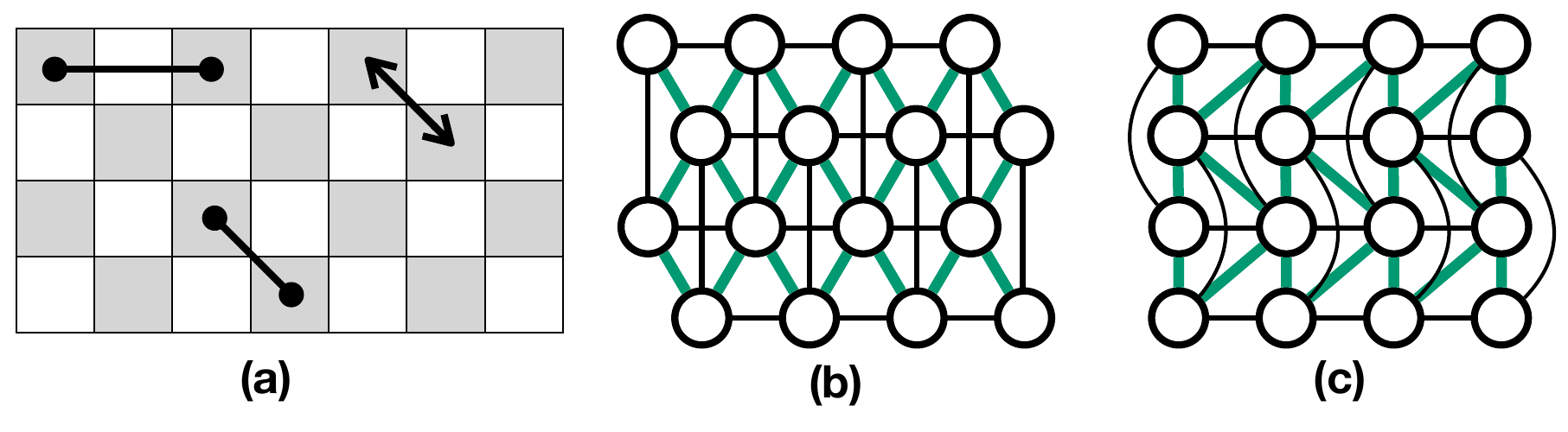}
    \caption{Lattice surgery mode: (a) Each tile represents a logical qubit. A grey tile is a data qubit. A white tile is an ancilla qubit. (b) The graph representing of qubits and their links. Green links are for faster SWAP. Black links are for slower SWAP. (c) A stretched grid of (b). }
    \label{fig:lattice-surgery}
\end{figure}

The surface code is a promising method for implementing fault-tolerant quantum computing (FTQC) as it has a high tolerance error threshold \cite{fowler+:physreva12, hua+:micro21}. The lattice surgery \cite{horsman+:njp12} is one of the modes of the surface code FTQC. In the lattice surgery mode, a 2D grid of logical qubits is formed by tiling the plane with rectangular tiles, as shown in Fig. \ref{fig:lattice-surgery}, where the grey tiles are logical computation qubits, and the white tiles are logical ancilla qubits. The ancilla qubits are necessary for implementing CNOT gates. Hence the ancilla qubits are interleaved with the data qubits. There are different ways to arrange ancilla qubits on the grid. The layout in Fig. \ref{fig:lattice-surgery} is a compact way for better resource usage. 

For long-range CNOT gates, SWAP gates are necessary \cite{beverland+:prx22}. However, unlike the NISQ devices, where the SWAP has the same latency across all links, the SWAP has different latencies on different links. SWAPs between diagonal (grey) tiles are faster, they have depth of two by using two ancilla qubits at once. SWAPs on horizontal or vertical tiles have to be implemented using three CNOT gates, which have a depth of six, as each CNOT has a depth of two \cite{beverland+:prx22}. Note that a CNOT can happen also between two diagonal tiles, with the same latency as happening between horizontal or vertical tiles. Such links are illustrated in Fig. \ref{fig:lattice-surgery} (a). In Fig. \ref{fig:lattice-surgery} (b), we describe the links between qubits using a graph, where each node is a data qubit. Fig. \ref{fig:lattice-surgery} (c) is a stretched representation of Fig. \ref{fig:lattice-surgery}, facilitating our further technical description.

\textbf{Discussion:} It is worth noting that no existing SWAP insertion approaches take the heterogeneous nature of the lattice surgery links for CNOT gates. The existing greedy SWAP insertion approaches such as Qiskit Sabre \cite{li+:asplos19} assume each link has the same latency, but it is not true for lattice surgery. The existing analytical approach, the LNN approach we introduced, does not take the heterogeneous links into consideration either. While it is possible to find a Hamiltonian path, in the graph in Fig. \ref{fig:lattice-surgery}, as the links have different latency, our approach significantly outperforms the LNN approach, as will be demonstrated in Section \ref{sec:eval}.

\subsection{Program Synthesis}
A program synthesis engine takes as an input a specification and an implementation. The specification describes what needs to be achieved, for instance, we require all CPHASE and H gates in QFT to be executed with respect to its dependencies in the transformed circuit. The implementation specifies the shape of the code that can potentially achieve the goal in the specification. However, such a code shape is roughly specified and certain parameters remain to be solved by the synthesizer. We leverage the regularity of the hardware, and represent the QFK kernel solutions as affine loops. Then we synthesize the loops using SKETCH \cite{sketch}, for certain scenarios, in our sub-kernel partitioning method.  


\section{Our Key Insights}
\label{sec:keyinsights}

We describe how we break the dependences in Section \ref{sec:breakdept}, perform sub-kernel partition in Section \ref{sec:recurQFT}, and one general scenario to combine the two in Section \ref{sec:unify}. 

\subsection{Breaking the Dependence} \label{sec:breakdept}

The original QFT kernel (described in Fig. \ref{fig:qftlogical} ) has two types of dependence.
\begin{itemize}
    \item \textbf{Type I dependence (has room to relax):} If two gates share the same control (or target), the one with a larger target (or control) index must run after the one with a smaller index. Assuming $G1 = G(q_i, q_j)$ and $G2 = G(q_i, q_k)$, if $j < k$ and $i \neq j$, G2 must run after G1. Same for the other way around for $G1 = G(q_j, q_i)$ and $G2 = G(q_k, q_i)$. 
    \item \textbf{Type II dependence (cannot relax any more):} If one gate's control is another gate's target, the latter must run after the former. For instance, if $G1 = G(q_i, q_j)$, $G2=G(q_j, q_k)$ where $i \neq j$ or $j \neq k$, G1 must run before G2.
\end{itemize}

There is a special case where a single-qubit Hadamard gate (H gate) also appears in the circuit. It still fits into the dependence constraint. We let H gate be presented as $G(q_i, q_i)$, allowing the control and the target to be identical in gate. Following the constraints defined above, it still works and defines the whole circuit's dependence.

However, we show that \emph{Type I dependence} constraint is unnecessary. Two CPHASE gates sharing the same qubit (either control or target) can commute \cite{alam+:dac20, lao+:isca22}. This result holds since the CPHASE gate is a diagonal unitary. One may wonder why Type II dependence does hold, as it doesn't matter whether two CPHASE gates share the control or target. The reason is that, we have H gate in between the CPHASE gates. The H gate does not commute with the CPHASE gate. Hence, from  $G1 = G(q_i, q_j)$ to $G2=G(q_j, q_k)$ where $i \neq j$, there is always one $G(q_j, q_j)$ between them. But for gates that share the control or target qubit, they can always appear in a row, in the original QFT. 

\begin{figure}[htb]
    \centering
    \includegraphics[width=0.48\textwidth]{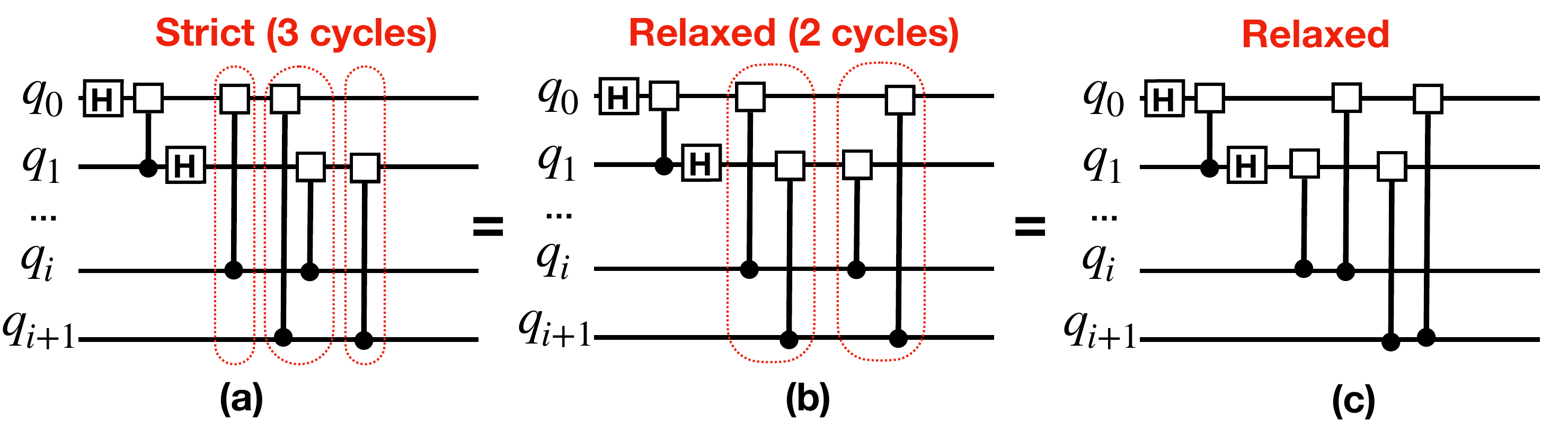}
    \caption{Benefits of relaxed ordering. (a) The original ordering of G($q_0$, $q_i$), G($q_1$, $q_i$), G($q_0$, $q_{i+1}$), G($q_1$, $q_{i+1}$). (b) (c) two other ways to reorder the gates}
    \label{fig:relax-ordering}
\end{figure}

\textbf{Examples for Breaking the Dependence} We show two scenarios for breaking such a dependence. Fig.~\ref{fig:relax-ordering} provides a simple example showing that breaking such dependency could help save running cycles, or provide a different ordering (that will be useful in the heavy-hex case later).  

\subsection{Sub-kernel Partitioning}
\label{sec:recurQFT}

We start with the 2-partition case, and then we extend it to the k-parition case for the QFT kernel. 

\textbf{\emph{2-partition QFT}}
First, we divide qubits into 2 subsets, and then divide the QFT process into three steps, without violating the dependence constraints (both type I and II).  

Let's assume we have qubits $U = \{q_0, ..., q_{N-1}\}$. $U$ is divided into two sub-groups $U1$ (consecutive qubits $q_0, q_1, ..., q_k$) and $U2$ ( consecutive qubits $q_{k+1}, ..., q_{N-1}$). We can prove that the following steps are correct.

\begin{itemize}
    \item Step 1: Execute QFT on $U1$.
    \item Step 2: Execute the gates between $U1$ and $U2$, by preversing their original order among themselves.
    \item Step 3: Execute QFT on $U2$.
\end{itemize}


\begin{figure}[htb]
    \centering
    \includegraphics[width=0.35\textwidth]{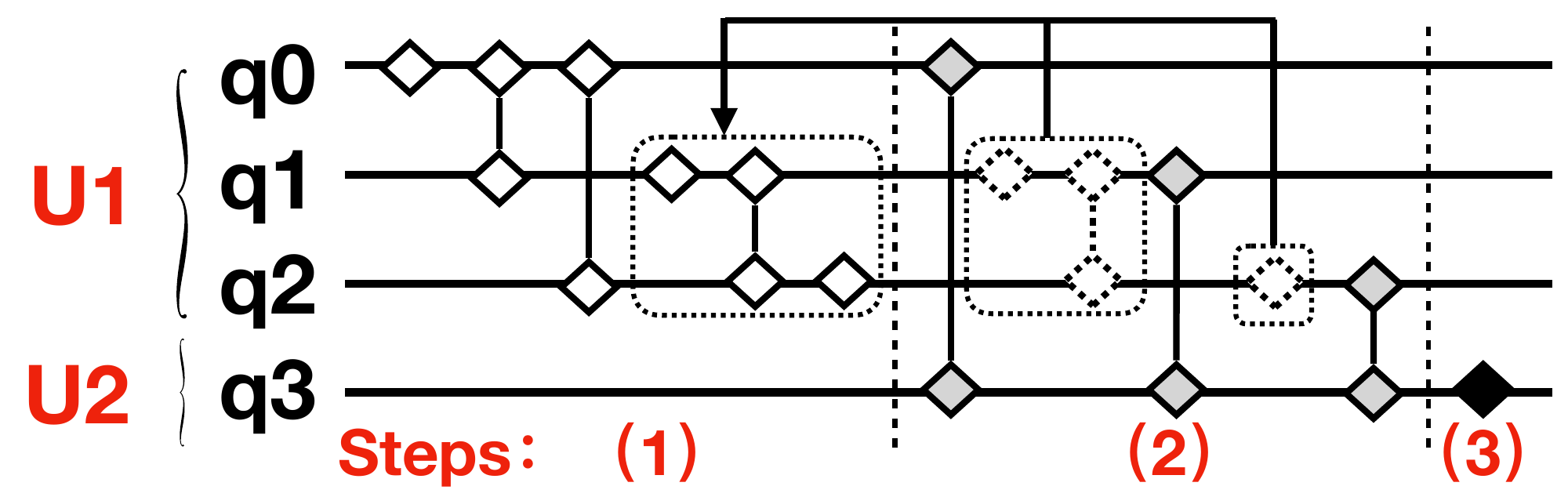}
    \caption{An example of dividing 4 QFT qubits into two subsets $U1$ and $U2$, and computation into thee steps. The dotted part shows the gates moved.}
    \label{fig:qft_Divide&Conquer}
\end{figure}
\begin{figure}[htb]
    \centering
    \includegraphics[width=0.4\textwidth]{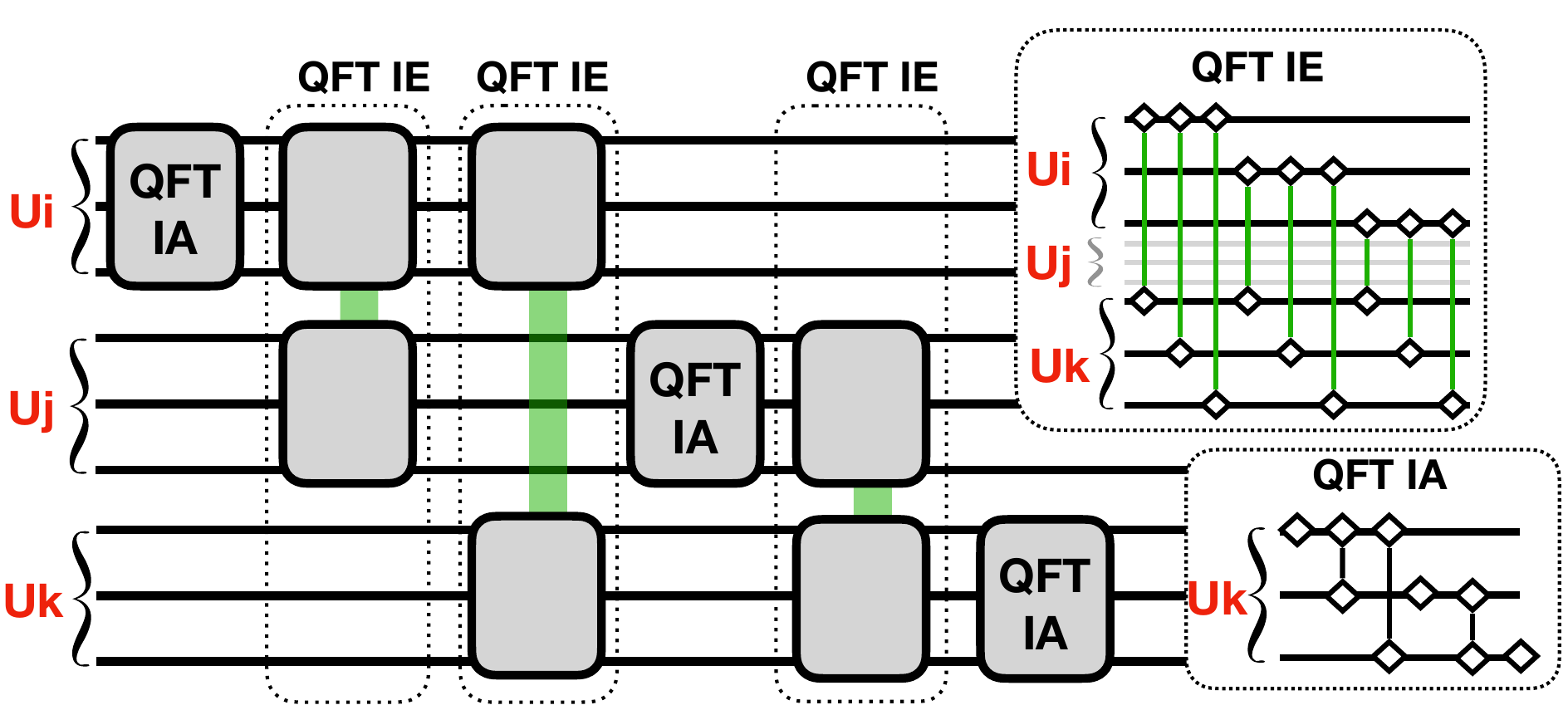}
    \caption{An example of the k-partition scheme for QFT. }
    \label{fig:qft_Div&Conq_K}

\end{figure}

 Due to the space limit, we sketch the proof for correctness. As long as a reordering satisfies these two types of dependence, it is valid. In the three steps above, we partition all gates involving qubits in $U1$ into two components: the interaction within $U1$ (Step 1), and the interaction between $U1$ and the rest of the qubits $U2$ (Step 2). Each component preserves its relative ordering before partition. We then first schedule all gates in Step 1 (within $U1$), all gates in Step 2 (between $U1$ and $U2$), and lastly Step 3 ( gates within $U2$).  An example is shown in Fig. \ref{fig:qft_Div&Conq_K}. 

Since it only has to satisfy Type II dependence as described in Section \ref{sec:breakdept}, we only need to prove Type II dependence is preserved.  

 For type II dependence, assuming $G1 = G(q_i, q_j)$, $G2=G(q_j, q_k)$, there are four cases for the location of $i$, $j$, and $k$. If all $i$, $j$, and $k$ are in $U1$, since we preserve the original relative order among these gates themselves, type II dependence is preserved. If $i$, $j$ are in $U1$, and $k$ in $U2$, G2 runs after G1, since G1 is in step 1, and G2 is in step 2. If $i$ in $U1$, and $j$, $k$ in $U2$, G1 still happens before G2, since G1 belongs to Step 2, and G3 belongs to step 3. If $i$, $j$, and $k$ are in $U2$, G1 happens before G2, as their relative ordering is preserved among these gates themselves in U2. Hence it is proved that such partition of QFT operations is correct. 

 In fact, it is trivial to see that Type I dependence is also preserved. We are simply regrouping gates that can run in parallel, as shown in the concrete example in Fig. \ref{fig:qft_Div&Conq_K}.

\textbf{\emph{$k$-partition QFT}}
In the previous example, we demonstrate the partition of QFT qubits into two sets, and the partition of QFT computation into three steps. It can in fact be extended to a k-partition QFT, where the qubits are divided into $k$ subsets $U_0, U_1, U_2, ...., U_{k-1}$. Again each subset needs not to have the same size. We can first divide the qubits into two subsets of $U_0$ and $U_{P2} = \{U_1, U_2, U_{k-1}\}$. Then we partition $U_{P2}$ into $U_2$ and $\{ U_3 ... U_{k-1}\}$. The whole process applies, and the proof of correctness still holds.

\begin{figure}[htb]
    \centering
    \includegraphics[width=0.4\textwidth]{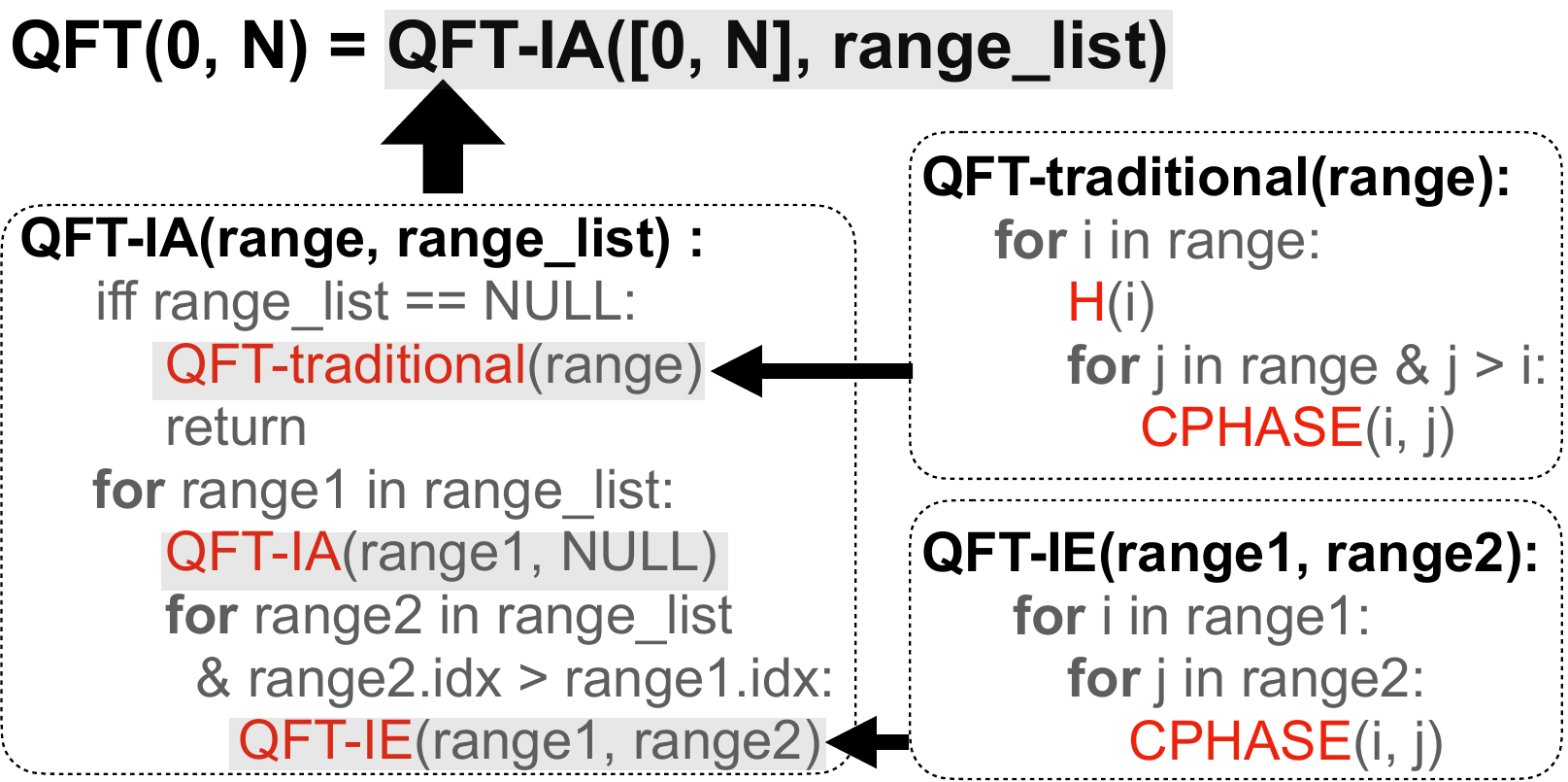}
    \caption{The k-partition QFT process. }
    \label{fig:DAC_Loop}
\end{figure}

We show the illustration of the k-partition QFT process in Fig. \ref{fig:qft_Div&Conq_K} and the pseudo-code in Fig. \ref{fig:DAC_Loop}. The function QFT-IA denotes QFT for a range of consecutive qubits and a list of small ranges.  The parameter $range\_list$ contains the list of small ranges. If the $range\_list$ is empty, QFT-IA degenerates to the traditional QFT operation, denoted as QFT-traditional, meaning we do not perform divide and conquer on this range of qubits. Otherwise, it performs a set of intra-QFT (QFT-IA) and inter-QFT (QFT-IE) on and between $U_i$, where $0 \leq i < range\_list.size()$. QFT-IE allows two small ranges of qubits to interact using CPHASE gates, and these two small ranges can have different sizes.

\subsection{Unifying the Two Insights}
\label{sec:unify}

We consider one possible scenario to unify these two insights. Since QFT-IE does not have any single-qubit gate (no gate $G(q_i, q_i)$), as it contains interaction between two sets, all gates in QFT-IE can commute. This corresponds to our discussion in Section \ref{sec:breakdept} where we can break Type I dependence. Hence, we combine Insight 1 and Insight 2. The idea is that, after breaking down the computation into QFT-IA and QFT-IE, we can optimize the QFT-IE component using commutativity. At the logical circuit level, it does not improve the circuit depth. But in the hardware circuit level, since we have to insert SWAPs, this flexibility can offer a 2X speedup. We discuss that in Section \ref{sec:sycamore}, \ref{sec:heavyhex} and \ref{sec:lattice}.  

We then have two different versions of QFT-IE. We denote the first version as \emph{QFT-IE-relaxed}, where gate reordering is exploited.  We refer to the second version as \emph{QFT-IE-strict}. Although QFT can directly use \emph{QFT-IE-relaxed}, we include the discussion for \emph{QFT-IE-strict} mainly because there are other circuits with similar structure to QFT but do not use CPHASE gates for two-qubit interaction.


The sub-kernel partition method QFT has two benefits. First, we can reduce a high-dimensional problem into low-dimensional problems recursively. Second, the unit size is broadly defined and each unit can have a different number of qubits. 

Since we can solve the LNN mapping problem, iff the units can be connected as if they are on ``a line", one can run the unit-based linear QFT as shown in Fig. \ref{fig:unitQFTline}. The trick is that it needs to ``SWAP" two adjacent units on the ``line", and perform QFT-IE between two adjacent units on a ``line", both in a hardware-compliant manner. For Google Sycamore and the lattice surgery grid, the units can form a ``line", as described in Section \ref{sec:sycamore} and \ref{sec:lattice}.




\section{Linear-depth QFT on Heavy-hex}
\label{sec:heavyhex}

\textbf{\emph{From Heavy-hex to Our Simplified Coupling Graph.}}
We generate a coupling graph for the heavy-hex architecture by deleting some connection lines from the original heavy-hex architecture (details in Appendix~1).
In the coupling graph (Fig. \ref{fig:initial_mapping_heavy_hex_dangling}), those qubits with only one qubit as its neighbor are defined as \textbf{dangling points} (except the first qubit), and all other qubits are defined to be the \textbf{main line} points.

\begin{figure}[htb]
    \centering
    \includegraphics[width=0.3\textwidth]{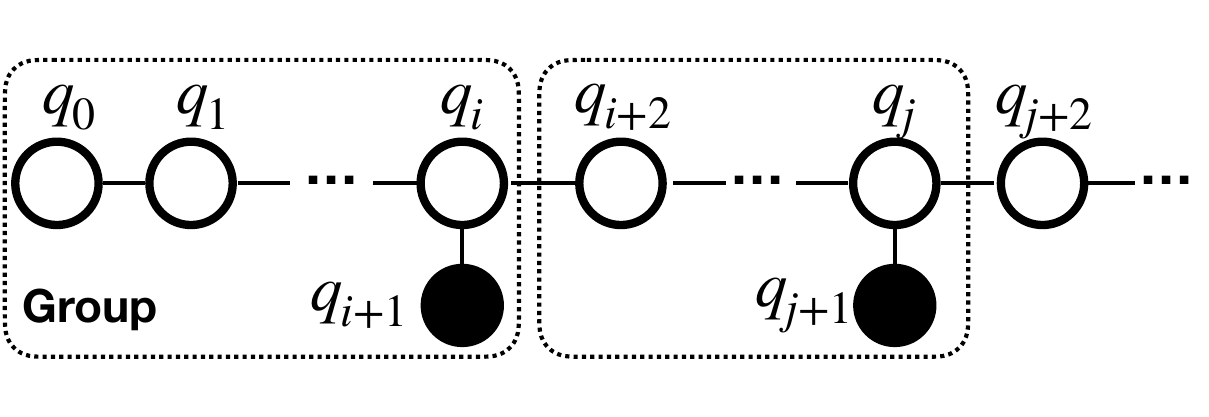}
    \caption{Initial mapping for Heavy-hex with dangling qubits (in black background): any node $i$ with a node below has an index $i+1$, and the right node has an index $i+2$. }
    \label{fig:initial_mapping_heavy_hex_dangling}
\end{figure}

\textbf{\emph{Initial Mapping in our Coupling Graph.}}
We put logical qubits' initial mapping in Fig.~\ref{fig:initial_mapping_heavy_hex_dangling}.
It follows a common rule: when we compare any two adjacent qubits, the qubit in the right/down position has a larger index.

\textbf{\emph{Intuition behind our approach. }}
Our method is a \textbf{non-trivial extension} of the LNN solution. We provide an in-depth understanding of the LNN solution first. 
\begin{enumerate}
    \item The displacement of all logical qubits follows an LNN pattern, characterized by a directional movement. Initially, all logical qubits shift towards the left. Upon reaching the leftmost position, they alter their trajectory.
    \item Each logical qubit only starts moving after being involved in one CPHASE gate operation. We denote that as the \emph{active} status of the qubit. A qubit will stop moving at one point, and we denote that as the \emph{in-active} status of the qubit.  
\end{enumerate}

For the simplicity of description, we specifically let $q_i$ denote the qubit connecting to the first dangling point, and $q_j$ denote the qubit connecting to the second dangling point, and respectively $q_{i+1}$ and $q_{j+1}$ are first two dangling points. 

Leveraging these findings, we describe our high-level idea below. We run the LNN operations on the first $i$ qubits until we position the smallest-index active qubit on the main line adjacent to the dangling point, initially $q_{0}$. Then we move the smallest-index active qubit on the main line to the dangling position by a SWAP. The process iterates along the main line, moving the smallest-index active qubit once it reaches the next dangling point by a SWAP
, and repeats. This sequence repositions first $L$ smaller-index qubits to dangling positions (where $L$ is the number of dangling points), disengaging them from LNN movement in the main line. 
The remaining qubits in the main line can continue to perform LNN QFT. We also stop every time after a SWAP layer and perform CPHASE cases, just like in the original QFT. Extra stops are needed for CPHASE between dangling points and their adjacent nodes. 

\begin{figure}[htb]
    \centering
    \includegraphics[width=0.49\textwidth]{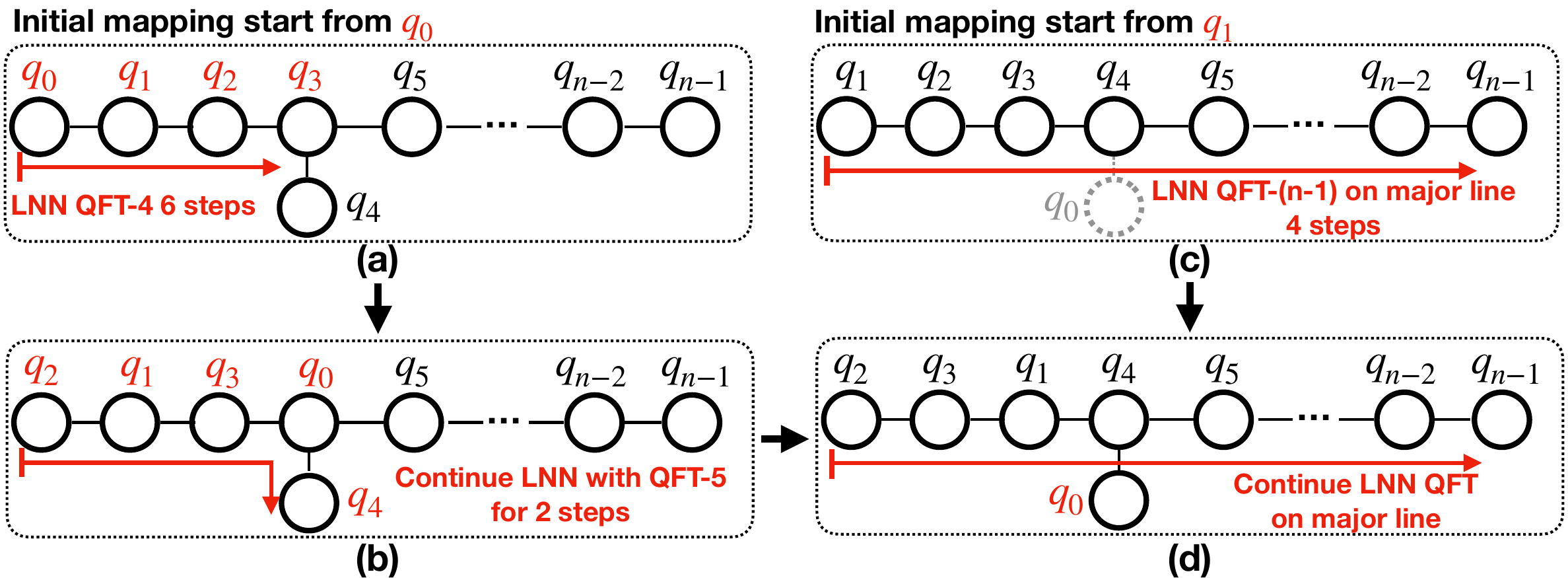}
    \caption{An example of running QFT in Heavy-hex with one dangling point. (a) Running LNN pattern in the main line with four active qubits. (b) Continue LNN pattern with the dangling qubit q4, which is identical to the LNN QFT-5 pattern.
    (c) Applying the LNN pattern on the main line with the first qubit $q_{1}$. (d) With dangling qubits, the qubit mapping would be the same as the case w/o dangling point.  
    QFT-K means the QFT with k qubits.  }
    
    
    \label{fig:swap_then_control_heavy_hex}
\end{figure}

Fig.~\ref{fig:swap_then_control_heavy_hex} provides a concrete example with only one dangling point. 
Before $q_{0}$ arrives the position above $q_{4}$, all qubits follows the linear QFT solution; 
when $q_{0}$ is above $q_{4}$, we implement one SWAP gate between them; 
then, 
the main line is an intermediate step of an LNN QFT for all nodes from $q_1$ to $q_{n-1}$ ((c) of Fig.~\ref{fig:swap_then_control_heavy_hex}).
The main line can continue performing gate operations using LNN QFT solution.

\textit{How do we ensure it also performs CPHASE between $q_0$ and qubits from $q_{5}$ to $q_{n-1}$?} In the straight line QFT, as shown in prior work \cite{zhang+:asplos21, maslov+:physreva07}, each qubit $q_i$ moves towards the left first, stops at the border for one step, and then to the right. 
All nodes after $q_{5}$ will move to the left end and then toward the right. 
When these nodes move to the left, they will be above $q_0$ at one point, where we let $q_0$ perform a CPHASE with these qubits. 
The interaction between $q_0$ and the nodes from $q_{5}$ to $q_{n-1}$ are completed. 


If there is more than one dangling point, we recursively handle this. We sketch the idea as follows. Specifically, it is as if we reduce one dangling point each time. Let's assume we have two nodes dangling from the main line (e.g., $q_{i+1}$ and $q_{j+1}$ in Fig. \ref{fig:initial_mapping_heavy_hex_dangling}). We follow the same solution as the single-dangling-point case for a few steps first.
The first time we swap $q_0$ with $q_{i+1}$, it is as if we can remove $q_0$ from the entire line, and then it reduces to the problem of having only one dangling point. We continue the solution for a straight line with one dangling point.
Hence, we can extend our approach to two dangling points and, subsequently, to $k$ dangling points. 
We put the details in Alg. \ref{heavyhex_qft}, and the final mapping of N qubits after the process is in Fig. \ref{fig:final_mapping_heavy_hex} in Appendix. 

\textbf{\emph{Breaking the Dependence.}} In this method, we leveraged the insight of breaking the dependence in Section \ref{sec:breakdept}. In the original QFT, $q_{0}$ interacts with all qubits $q_2$ to $q_{last}$ before $q_{1}$ interacts with all of them. However, in our method, qubit $q_0$ is placed in the dangling point $q_{i+1}$. And qubit $q_{1}$ will be placed in the second dangling position of $q_{j+1}$. Hence, $q_1$ interacts with the qubits of indices larger than $j+2$ before $q_{0}$ interacts with them. This is okay, our Type II dependence only requires G($q_i$, $q_j$) to happen before G($q_j$, $q_k$). Since G($q_0$, $q_1$) already has happened,  for any qubit $q_k$, where $k \geq j+2$, G($q_1$, $q_k$) can happen before G($q_0$, $q_k$).  An example is shown in Fig. \ref{fig:relax-ordering}(c).   




\begin{algorithm}
	\caption{Heavy-hex qubit mapping procedure}           \label{heavyhex_qft}
        \scriptsize
	\begin{algorithmic}
            \For {$i=0;$ $i<N_{1};$ i++}
            \If{$Q[0][i] < Q[0][i + 1] \wedge C(Q[0][i], Q[0][i + 1]) == 0$} 
                \State $C(Q[0][i], Q[0][i + 1]) \gets 1;$ i++;\algorithmiccomment{CPHASE to the right qubit}
            \EndIf 
            \If{$Q[0][i] < Q[0][i + 1] \wedge C(Q[0][i], Q[0][i + 1]) == 1$} 
                \State $SWAP(Q[0][i], Q[0][i + 1]);$ i++;
                \algorithmiccomment{SWAP with the right qubit}
            \EndIf 
            \If{$Q[0][i] < Q[1][i] \wedge C(Q[0][i], Q[1][i]) == 0$} 
                \State $C(Q[0][i], Q[1][i]) \gets 1$
                \algorithmiccomment{CPHASE to the qubit below}
            \EndIf
            \If{$Q[0][i] > Q[1][i] \wedge C(Q[1][i], Q[0][i]) == 0$} 
                \State $C(Q[1][i], Q[0][i]) \gets 1$
                \algorithmiccomment{CPHASE to the qubit below}
            \EndIf
            \If{$Q[0][i] < Q[1][i] \wedge C(Q[0][i], Q[0][i + 1]) == 1$} 
                \State $SWAP(Q[0][i], Q[0][i+1]) \gets 1$
                \algorithmiccomment{SWAP with the qubit below}
            \EndIf 
            \EndFor
	\end{algorithmic}
\end{algorithm}


\textbf{\emph{Time Complexity for Special Case.}}
We calculate the time complexity for a special case with a dangling point for every four qubits on the main line. This is analogue to the heavy-hex case. 
In this case, each group consists of 5 qubits.
80$\%$ of qubits are in the main line while the remaining 20$\%$ are in the dangling position. 
The distance between 2 adjacent dangling points is 4. 
Our proof (in Appendix~2) shows that an extra 25 steps are required for one added 5-qubit group. Therefore, the final complexity time is $5N +O(1)$, where N is the number of qubits in the heavy-hex architecture. 



\textbf{\emph{Time Complexity Bound for General Case.}} 
Additionally, we also provide an upper bound for a general case where there are N1 qubits in the main line and N2 dangling qubits. The distance between 2 nearby dangling nodes might be different. The worst case is that every time there is a SWAP operation in the main line, another cycle is needed for the CPHASE operation with dangling points. The final time complexity upper bound is $6N + O(1)$. Detailed proof is in Appendix~3.

\section{Linear-depth QFT on Google Sycamore}
\label{sec:sycamore}

\begin{figure}[htb]
    \centering
    \includegraphics[width=0.35\textwidth]{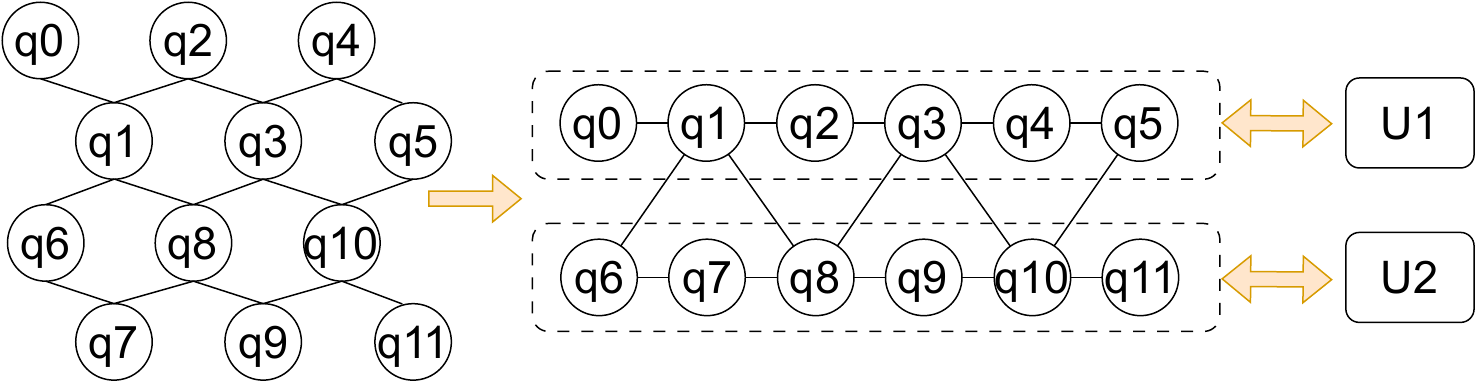}
    \caption{Unit definition in Sycamore architecture. }
    \label{fig:googlesycamore_unit}
\end{figure}

\textbf{\emph{Unit definition}}
We combine every two rows together as a unit shown in Fig.~\ref{fig:googlesycamore_unit}.
There are $m/2$ units in one $m * m$ Google Sycamore, each containing $2*m$ qubits.

\textbf{\emph{Unit SWAP.}}
This design makes it possible to complete the unit swap between 2 adjacent units in 3 steps. We describe it for the particular simple example in Fig. \ref{fig:googlesycamore_unit}. 
{
\small
\begin{verbatim}
    parallelSWAP ({q1, q3, q5} {q6, q8, q10})
    parallelSWAP ({q1, q3, q5} {q7, q9, q11}) 
                 ({q0, q2, q4} {q6, q8, q10})
    parallelSWAP ({q0, q2, q4} {q7, q9, q11})
\end{verbatim}
}



We decompose the qubit mapping problem of QFT over the Sycamore architecture into two categories: intra-unit and inter-unit mapping.

\textbf{\emph{Intra-unit QFT.}} Given all qubits within one unit are in a line, we leverage the mapping algorithm used for QFT over LNN architecture for intra-unit qubit mapping.

\begin{figure}[htb]
    \centering
\includegraphics[width=0.35\textwidth]{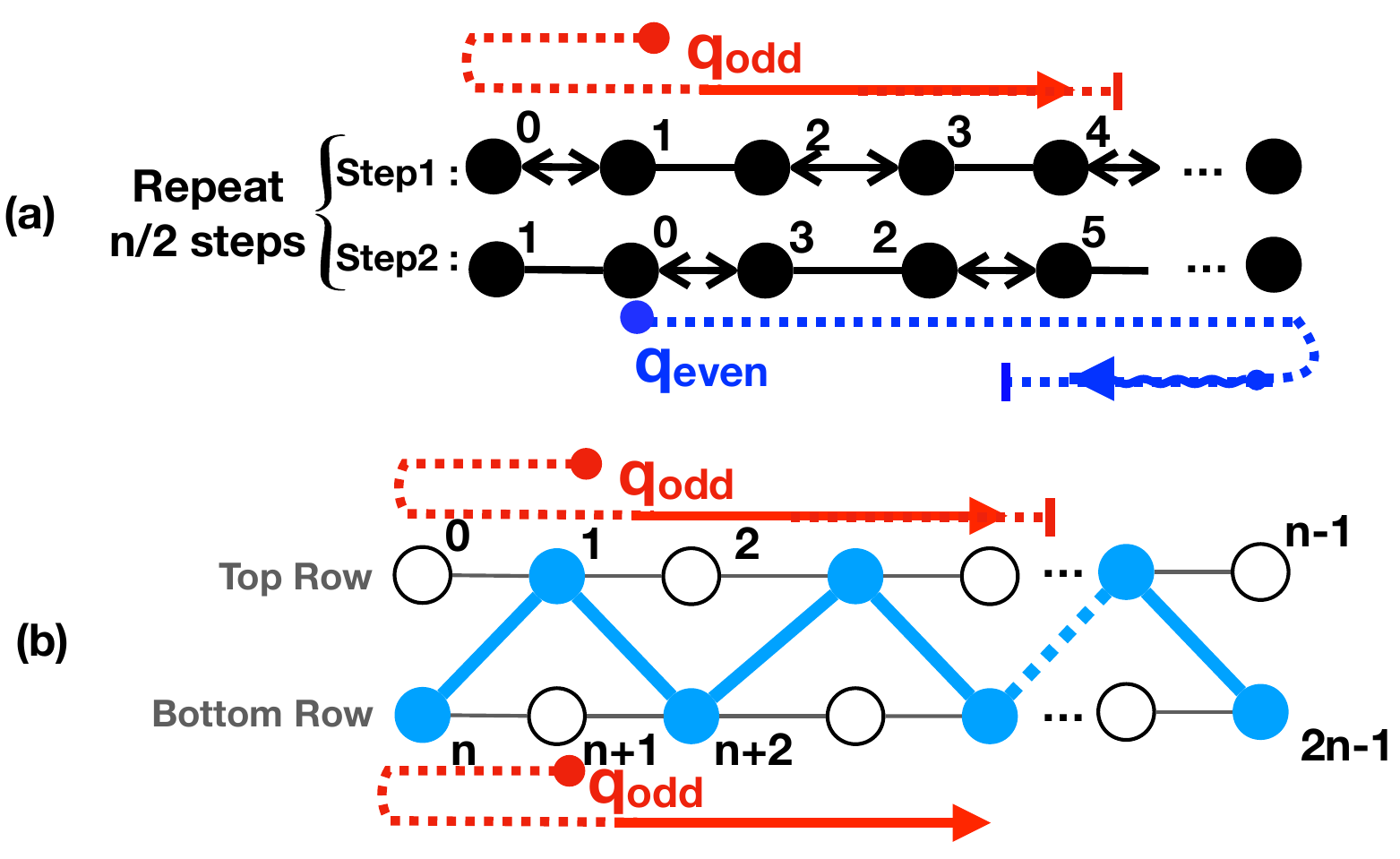}
    \caption{Travel paths for inter-unit QFT in Sycamore architecture. (a) Two consecutive SWAP layers for \textbf{one unit} over time. Each qubit has a different neighbor after one SWAP step. Both units (top and bottom) sync this way, although only one unit is drawn here.   
    (b) The top and bottom rows sync with the same travel path. Qubits in a unit have different neighbors at each step from the other unit, using the top-down links highlighted in blue. }
\label{fig:sycamorepath}
\end{figure}

\textbf{\emph{Inter-unit QFT (Relaxed Ordering).}} 
We apply the relaxed ordering in IE interactions between two adjacent units. 
We briefly show the qubit travel path in a LNN-inspired pattern (found by program synthesis) shown in  Fig.~\ref{fig:sycamorepath} (a) for both the top and bottom units in one adjacent pair for bipartite all-to-all interactions. Each unit follows the LNN-inspired pattern, so every node is a neighbor to the other nodes in the same unit. Since both units are synced, and there are diagonal links between two units, this also ensures each node in one unit is a neighbor to all nodes in another unit once, except to the node in \textit{the same column}. 

\textbf{Our approach.}
We discovered this inter-unit solution by parameterizing the relative LNN-inspired moving pattern in the top and bottom rows with the help of program synthesis (details in Appendix~5).
Specifically, we find that if we sync the top and bottom to follow the same travel path in Fig.~\ref{fig:sycamorepath}(a), an inter-unit interaction pattern emerges. 
In Fig.~\ref{fig:sycamorepath}(b), there is a diagonal link between one qubit on the top row and the qubit on the bottom row \emph{if the two qubits' column index differ by 1.} Note that the moving pattern in Fig. \ref{fig:sycamorepath} (a) is similar to that in LNN, \emph{but not the same}. For LNN, only on average half of the links are utilized, but in our pattern here, all links in a perfect matching are used. 

Following the same travel path can guarantee that all top-row and bottom-row qubit pairs will be connected through the inter-unit link at least once, except for the pairs in the same column.
Fortunately, figuring out a solution for these missing CPHASE gates is easy. 
We can SWAP one of them horizontally with its neighbor at the top (bottom) row, keep the other one on the bottom (top) row unchanged, run a CPHASE gate, and then revert the qubit to its original location with the same SWAP.

The QFT-IE-relaxed version is two times faster than the QFT-IE-strict version. A detailed solution is below:
\begin{footnotesize}
\begin{verbatim}
    for (i = 0; i <= m; i += 1)
       CPHASE on all inter-unit connections   
      // Intra-unit swap
      beg = i mod 2
      intra_swap(beg_pos=beg, end_pos=m, UnitID=0)
      intra_swap(beg_pos=beg, end_pos=m, UnitID=1)
\end{verbatim}
\end{footnotesize}


Another benefit of our solution is that both QFT-IE-strict and QFT-IE-relax will mirror the position of all qubits within a unit. It helps further processing that for QFT-IA. 

\begin{figure}[htb]
    \centering
    \includegraphics[width=0.4\textwidth]{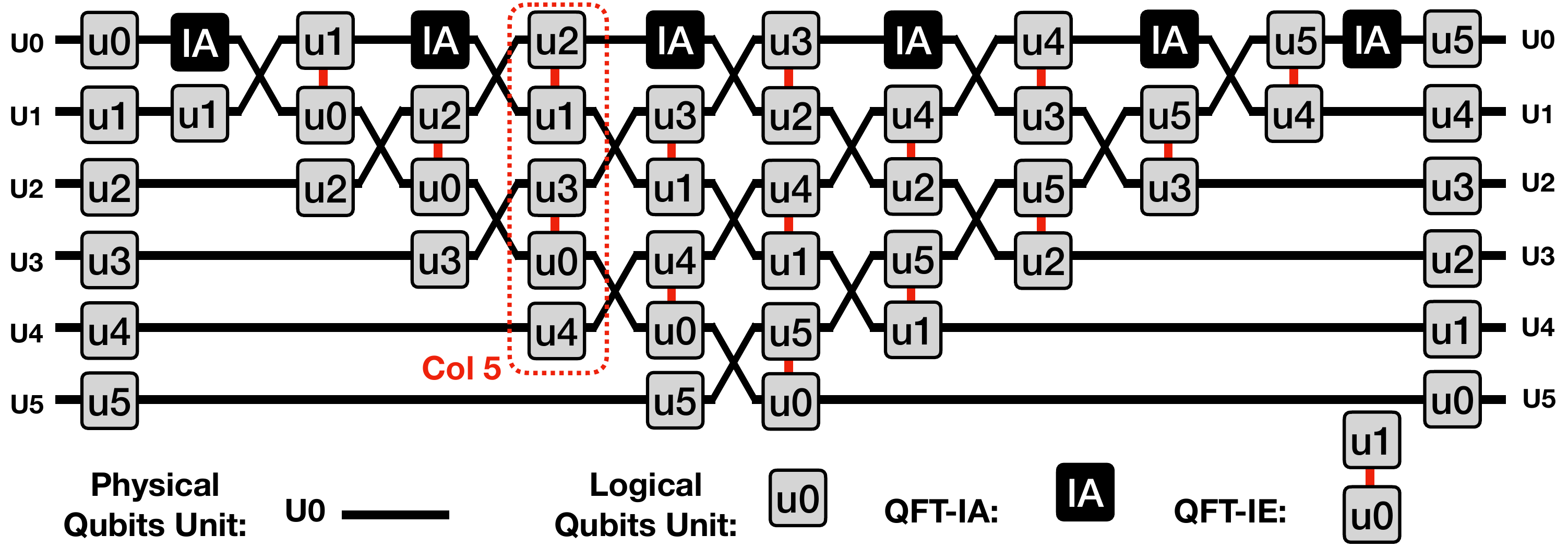}
    \caption{Unit-wise QFT using the recursive QFT scheme. This is analogous to the original LNN QFT solution. }
    \label{fig:unitQFTline}
\end{figure}

\textbf{\emph{Time Complexity}} 
Assuming $N=m*m$ Sycamore grid, where m is the original row size.  In our formulation, each unit consists of 2m qubits, and there are $\frac{m}{2}$ units. Each unit is connected as if they are on a line, as described earlier. We will do the QFT-IA and QFT-IE using the divide-and-conquer method. The hardware-compliant unit QFT on LNN is presented in Fig. \ref{fig:unitQFTline}. 

\textbf{\emph{QFT-IE-relaxed}}
For the QFT-IE-relaxed case, each QFT-IE takes $3*(2m+1)$ time steps, and each QFT-IA takes $4*(2m)-6$ time steps. 
Each unit SWAP takes three-time steps. There are in total $(m/2)+O(1)$ QFT-IE parallel steps, and in total $(m/2)+O(1)$ (QFT-IE, QFT-IA) mixed steps. There are a total of $2*(m/2)-3$ unit SWAP steps. Hence, the total time complexity is $7m^2+O(m) = 7N + O(\sqrt(N))$, where N is the total number of qubits.


\section{Linear-depth QFT on FT Backend}
\label{sec:lattice}


\begin{figure}[htb]
    \centering
    \includegraphics[width=0.5\textwidth]{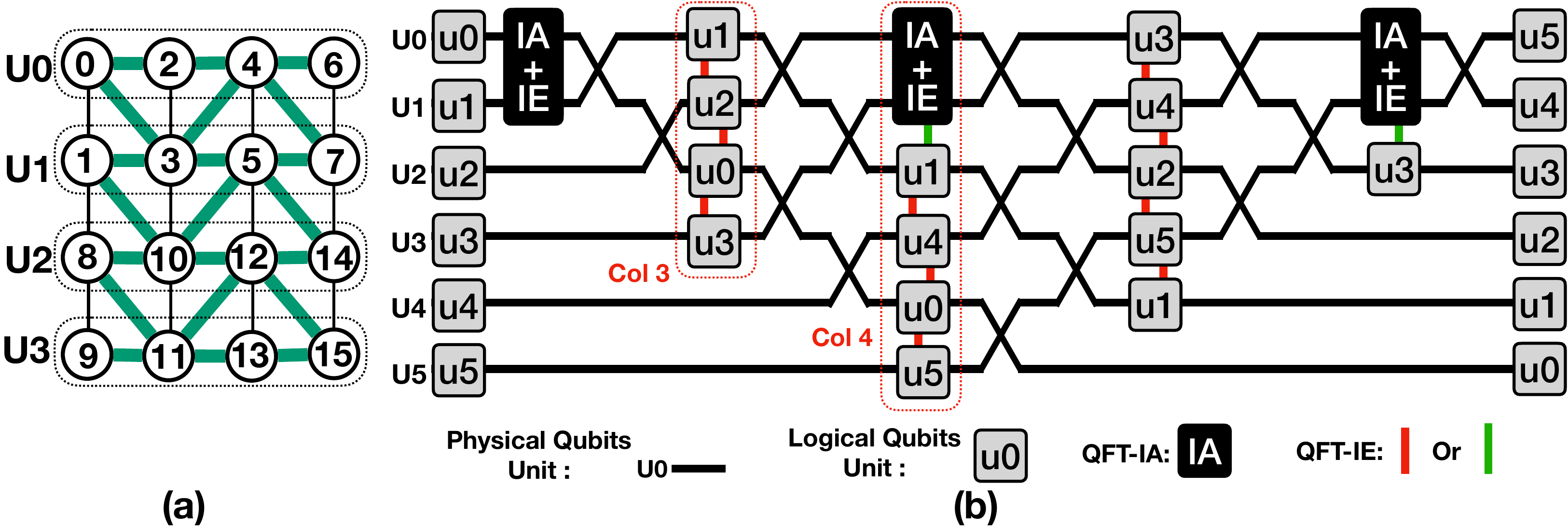}
    \caption{(a) Initial Mapping in Lattice Surgery. The layout of FT architecture is rotated to make all SWAP edges horizontal. Certain edges are unused and have been eliminated. (b) Optimized Unit Movement. }
    \label{fig:optimizedUnitMove}
\end{figure}

\textbf{\emph{Unit Definition and Movement}}
 We consider each row in the (rotated) FT grid as a unit and place qubits in natural number ordering from left to right in a zigzag way for every two units ( Fig.\ref{fig:optimizedUnitMove}(a) ).
 Similarly, we divide the problem into three sub-problems: intra-unit interaction QFT-IA, inter-unit interaction QFT-IE, and unit swapping.

We optimize the unit movement for the FT grid architecture as shown in the example in Fig. \ref{fig:optimizedUnitMove}(b). We run every \textit{two} parallel unit-SWAP layers in a row, before performing the inter-unit CPHASE operations, instead of having one unit-swap layer and one CPHASE layer interleaved.  The unit swap is trivial by applying transversal SWAPs using the vertical links between two units in one step. Those vertical links are CNOT-only links, so one vertical SWAP costs 3 CNOT gates. 
This choice could save SWAP costs for inter-unit interactions, which will be discussed later.

\textbf{\emph{Intra and inter-unit QFT: IA+IE.}} In our updated unit movement,  every disjoint pair of units, ($U_{i}$, $U_{i+1}$), where $i$ is an even index, would appear at the top two rows just once. Then, we can apply the 2$\times$N QFT pattern, introduced in \cite{zhang+:asplos21} to complete mixed intra- and inter- unit operations. For instance, the unit pairs ($U_{0}$, $U_{1}$), ($U_{2}$, $U_{3}$), and ($U_{4}$, $U_{5}$) depicted in Fig. \ref{fig:optimizedUnitMove} (b), which are enclosed in black boxes, execute the 2$\times$N QFT pattern. 


\begin{figure}[htb]
    \centering
    \includegraphics[width=0.35\textwidth]{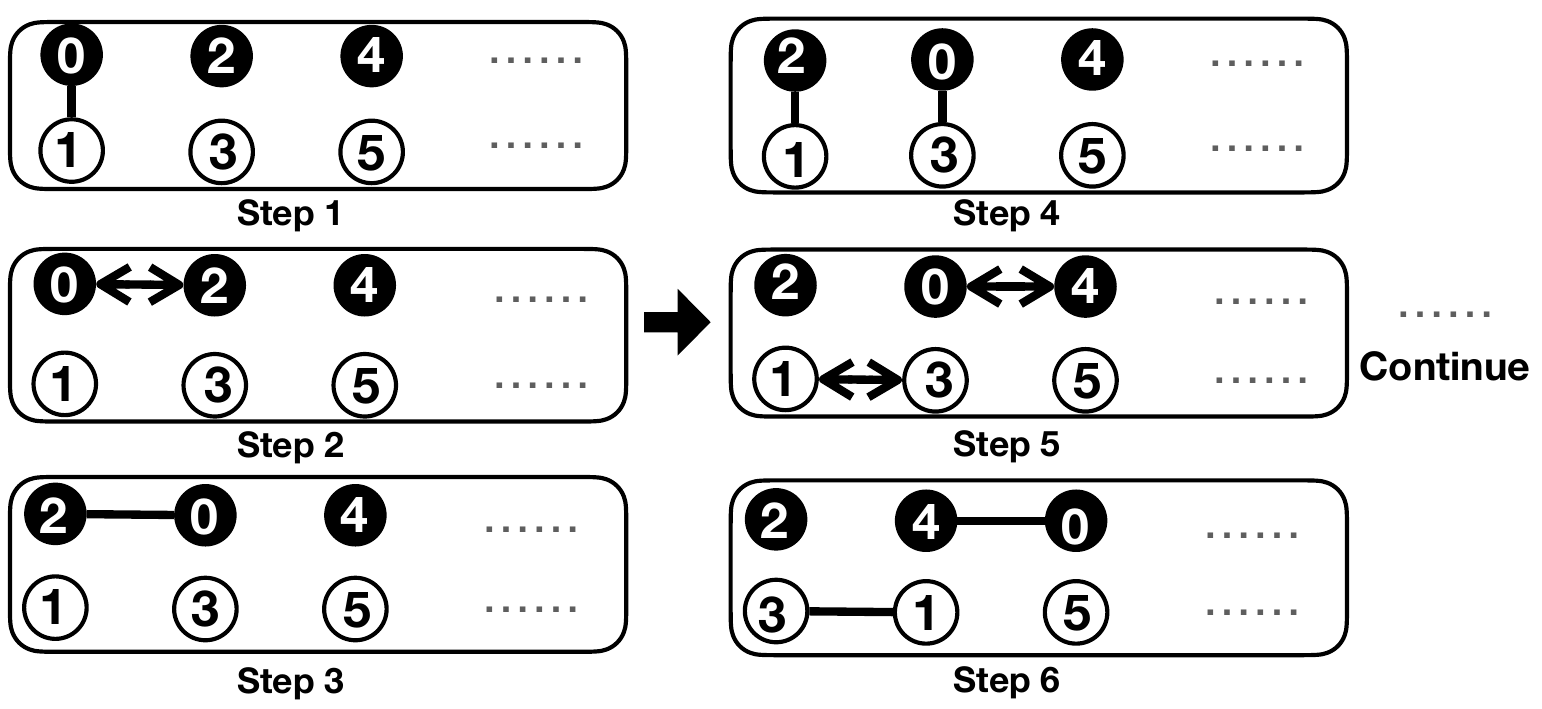}
    \caption{The first six steps of the 2$\times$N QFT pattern.}
    \label{fig:2xN_pattern}
\end{figure}

\textbf{The intuition for 2$\times$N QFT}
is to make the SWAP gates for intra-unit interactions benefit inter-unit interactions. Due to the gate dependency, it is as if we apply the LNN pattern on both the top and bottom units, but the bottom unit starts one step late. If the bottom unit does not start one step late, each qubit on the top unit always has the same (column) neighbor from the bottom unit, preventing full inter-unit operations. The initial mapping is designed to obey gate dependency. 
The 2$\times$N QFT pattern contains three repeated steps: inter-unit interaction, intra-unit SWAP, and intra-unit interactions. 
We show a part of the 2$\times$N pattern in Fig. \ref{fig:2xN_pattern}.

\textbf{The inter-unit interactions (pure IE -- red links in figures)}  follow the QFT-IE-Relaxed strategy. We let each qubit in the top unit interact with every qubit in the bottom unit. The qubit movement within each unit is described in Fig. \ref{fig:sycamorepath} (a). Again, the bottom unit starts one step late to prevent a qubit from always having the same neighbor qubit in the same column. We can achieve bipartite all-to-all inter-unit interactions with N steps of qubit movement and N steps of qubit interactions (this is developed via our program synthesis approach, details in Appendix Section 7).

\textbf{Saving SWAP costs for inter-unit interactions.} Since there is no gate dependency in the pure IE interactions like Col3 in Fig. \ref{fig:optimizedUnitMove}(b). The compilation for this part is the same as compiling a  QAOA circuit with bipartite all-to-all interactions in two units. We can adapt the SWAP saving strategy from \cite{jin_asplos23_qaoa}.

Before our optimization, there is no inter-unit interactions between two consecutive pairs of units (the last unit of the first pair and the first unit of the second pair). For example, there is no red link between two units $U_2$ and $U_0$ in the fifth column of Fig. \ref{fig:unitQFTline}. But now, we have a link between unit $U_{2}$ and $U_{0}$ in the third column of Fig. \ref{fig:optimizedUnitMove}(b). If we apply the same pattern to two pairs of units ($U_{1}$, $U_{2}$) and ($U_{0}$, $U_{3}$) in the third column of Fig. \ref{fig:optimizedUnitMove}(b), then, the trajectory of qubits within  $U_{1}$ is the same as the trajectory of qubits within  $U_{0}$.  Such that, if we can apply CPHASE gate between qubits in ($U_{1}$, $U_{2}$), we could also apply CPHASE gate between qubits in ($U_{2}$, $U_{0}$).

 One special case of inter-unit interactions is highlighted in green in the fourth column of Fig. \ref{fig:optimizedUnitMove}(b). Those inter-unit interactions involve one unit that is doing mixed intra- and inter-unit interactions. We pause the 2$\times$N pattern on the top two units after or before each SWAP step in Fig. \ref{fig:2xN_pattern}, letting the qubit on the physical $U1$ interact with qubits on the physical $U2$. The all-to-all interactions are still guaranteed since the qubit movement in the relaxed ordering is similar to the qubit movement in the 2$\times$N QFT pattern with different start times.

\textbf{\emph{Time Complexity.}} As for the time complexity, we use unit QFT on a line, as shown in Fig. \ref{fig:optimizedUnitMove}(a). Assuming we have $N=m*m$ qubits. Each unit has $m$ qubits. 

\textbf{\emph{QFT-IE-Relaxed}} In this case, we do not consider strict ordering in QFT-IE. The time complexity of QFT-IE without QFT-IA is $3m+O(1)$. It happens $m/2 - 1$ times. The time complexity of QFT-IE is $1.5m^{2}+O(m)$. Each mixed QFT-IE QFT-IA has complexity $6m+O(1)$. The detail of this time complexity refer to \cite{zhang+:asplos21}. Due to the relax reordering, we pause  mixed (QFT-IA, QFT-IE) for m steps for the inter-unit interactions between the mixed part and the unit below the mixed part. Mixed (QFT-IA, QFT-IE) steps has $7m+O(1)$ depth, and it happens $m/2$ times. For this part, the time complexity is $3.5m^{2} +O(1)$. In total, the time complexity is $5m^{2}+O(1)=5N+O(1)$.

\section{Evaluation}
\label{sec:eval}
We evaluate our 
QFT mapping 
approach against other compilation approaches over 
several 
hardware architectures, including NISQ architecture and FT architecture, in several aspects. Specifically, we want to ask 2 major questions. \textbf{Q1:} How long does it take to find the solution? 
\textbf{Q2:} What is the quality of these generated solutions?

We write an open-source simulator~\cite{anonymous_github} to check 
the correctness of our outcome. We measure the quality of the compilation outcome, mainly based on circuit depth and gate count. Due to the noisy feature of quantum gate operations, smaller depth and fewer gate operations mean a lower possibility of being affected by external noise.
We present the compilation time in seconds.
If it cannot complete in 2 hours, we refer to it as ``time-out" in Table~\ref{table:expr}. 

Finally, we explore the potential for applying our approach to larger-scale QFT circuits utilizing a fault-tolerant backend. It would be useful for us to exploit regularity in both a small-scale NISQ architecture and a large-scale FT architecture. 

\textbf{\emph{Baseline and benchmark selection.}} 
As for the baseline, we compare our approach against three approaches: LNN~\cite{maslov+:physreva07}, SATMAP~\cite{molavi+:micro22} and SABRE~\cite{li+:asplos19}.
SATMAP searches over the whole space and output optimal solutions (with respect to gate count) at the cost of long-running time.  SABRE does qubit mapping using a series of heuristics, making it possible to quickly output the potentially suboptimal results. LNN~\cite{zhang+:asplos21} is only tested on Lattice Surgery because we cannot find a hamiltonian path on Sycamore and Heavy-hex architectures.
The benchmarks encompass various scales of QFT, each configured differently across a range of NISQ and FT architectures. 

\textbf{\emph{Diverse architecture backends.}} We use three types of architectures: Google Sycamore, 
heavy-hex, and lattice surgery, among which only lattice surgery is a FT architecture particularly suited for implementing larger QFT kernels, while others are NISQ architectures. 


Google Sycamore has m by m configuration, where the total number of qubits is $N=m*m$. We test configurations where m is an even number.  
For heavy-hex, we unroll it to a line with dangling points \cite{weidenfeller+:arxiv22qaoa}. 
Based on the description in Sec~\ref{sec:heavyhex}, there are $N/5$ groups and each group has 5 qubits. Among all these 5 qubits, 4 are in the main line and 1 sits as a dangling point. Therefore, we only test the heavy-hex architecture whose qubit number is a multiple of 5. 
For lattice surgery, the total number of qubits is $N=m*m$. Based on our approach in Sec~\ref{sec:lattice}, we only consider the case where $m$ is larger than 10.

\subsection{Quality of Compilation Outcome for Small-scale QFT kernel on NISQ Backends}

Due to the error rates, only small-scale QFT circuits are deemed suitable for execution on NISQ backends including Sycamore, and Heavy-hex.
Thus, in our experiments conducted on these architectures, we assess the quality of compilation outcome across varying numbers of qubits, up to a maximum of 100, by adjusting the value of $m$.
 
In this evaluation, we focus on two principal aspects.
Firstly, we examine the time each approach takes to output the compilation result, with a preference for faster speed.
Secondly, we assess the quality of the outcomes, which includes considerations of circuit depth and the count of gates required to complete the QFT. 
Achieving the QFT with a smaller depth and a fewer number of SWAP gates is considered advantageous. What needs to be mentioned is that our method 
does not have compilation time as
it is an analytical approach.

\subsubsection{Compilation time}
A portion of the experimental outcomes is presented in Table~\ref{table:expr}. SATMAP, as an optimal solver, delivers favorable results in terms of circuit depth and gate count. However, its search space grows exponentially, leading to prolonged solution times. For example, when setting a time-out limit of 2 hours, SATMAP fails to produce results in most cases when there are more than 10 qubits.

In comparison, SABRE achieves outcomes significantly faster than SATMAP. However, as the number of qubits increases in one configuration, the running time of SABRE increases as well (e.g., 55s for 30*30 lattice surgery). 

\begin{table}[]
\centering
\resizebox{0.48\textwidth}{!}{
\begin{tabular}{|l|l|ll|lll|lll|}
\toprule
\multirow{2}{*}{Architecture} & \multirow{2}{*}{\# qubits} & \multicolumn{2}{c|}{Our approach}    & \multicolumn{3}{c|}{SATMAP}                                        & \multicolumn{3}{c|}{SABRE}                                        \\ \cline{3-10} 
                      &                            & \multicolumn{1}{l|}{Depth} & \# SWAP & \multicolumn{1}{l|}{CT(s)}  & \multicolumn{1}{l|}{Depth} & \# SWAP & \multicolumn{1}{l|}{CT(s)} & \multicolumn{1}{l|}{Depth} & \# SWAP \\ \midrule
m*m Sycamore          & 2*2                        & \multicolumn{1}{l|}{10}    & 6       & \multicolumn{1}{l|}{1.75}   & \multicolumn{1}{l|}{10}    & 3       & \multicolumn{1}{l|}{0.28}  & \multicolumn{1}{l|}{11}    & 3       \\
m*m Sycamore          & 4*4                        & \multicolumn{1}{l|}{81}    & 116     & \multicolumn{1}{l|}{TLE}       & \multicolumn{1}{l|}{N/A}      & TLE     & \multicolumn{1}{l|}{0.29}  & \multicolumn{1}{l|}{102}   & 62      \\
m*m Sycamore          & 6*6                        & \multicolumn{1}{l|}{208}   & 540     & \multicolumn{1}{l|}{TLE}       & \multicolumn{1}{l|}{N/A}      & N/A     & \multicolumn{1}{l|}{0.46}  & \multicolumn{1}{l|}{363}   & 484     \\
\midrule
Heavy-hex             & 2*5                        & \multicolumn{1}{l|}{39}    &  40    & \multicolumn{1}{l|}{439.79} & \multicolumn{1}{l|}{44}    & 37      & \multicolumn{1}{l|}{0.30}  & \multicolumn{1}{l|}{43}    & 36      \\
Heavy-hex             & 4*5                        & \multicolumn{1}{l|}{89}      &   160      & \multicolumn{1}{l|}{TLE}       & \multicolumn{1}{l|}{N/A}      & N/A     & \multicolumn{1}{l|}{0.31}  & \multicolumn{1}{l|}{134}    & 196      \\
Heavy-hex             & 6*5                        & \multicolumn{1}{l|}{139}      &     360    & \multicolumn{1}{l|}{TLE}       & \multicolumn{1}{l|}{N/A}      & N/A     & \multicolumn{1}{l|}{0.56}  & \multicolumn{1}{l|}{229}   & 523    \\
\midrule
Lattice surgery             & 10*10                        & \multicolumn{1}{l|}{476}      &     2700    & \multicolumn{1}{l|}{TLE}       & \multicolumn{1}{l|}{N/A}      &  N/A    & \multicolumn{1}{l|}{0.38}  & \multicolumn{1}{l|}{981}   & 2365    \\
Lattice surgery              & 20*20                        & \multicolumn{1}{l|}{1961}      &     41880    & \multicolumn{1}{l|}{TLE}       & \multicolumn{1}{l|}{N/A}      &  N/A    & \multicolumn{1}{l|}{8.67}  & \multicolumn{1}{l|}{11818}   & 55474    \\
Lattice surgery              & 30*30                        & \multicolumn{1}{l|}{4446}      &     208800    & \multicolumn{1}{l|}{TLE}       & \multicolumn{1}{l|}{N/A}      &  N/A    & \multicolumn{1}{l|}{55.26}  & \multicolumn{1}{l|}{47161}   & 305807    \\
\bottomrule
\end{tabular}

}
\begin{small}
\caption{Our approach vs. SATMAP and SABRE arcoss different architecture (CT: compilation time, TLE: timeout after 2h).}
\label{table:expr}
\end{small}
\vspace{-10pt}
\end{table}


\vspace{-0.11in}
\subsubsection{Gate count and Depth}
Regarding the number of SWAP gates, the general trend indicates that SABRE requires more SWAP gates or larger depth than ours, except for configurations with a very small number of qubits. 
Compared with SABRE, our approach have up to $53\%$ fewer SWAP gate count and $92\%$ fewer depth. In the Sycamore backend with up to 100 qubits, our methodology yields a depth cost approximately 50\% lower than that of SABRE, alongside a 20\% reduction in the number of SWAP gates. For the Heavy-hex backend, our approach significantly reduces the depth cost to just 24\% of SABRE's and cuts the number of SWAP gates needed to 48\% of those required by SABRE.
It's noteworthy that SABRE's performance is not consistently stable; in certain instances, it may result in a smaller gate count and depth for larger QFT sizes, particularly in the heavy-hex architectures.

\textit{Generally, we discover that our approach produces significantly better outcomes for QFT qubit mapping in terms of circuit depth and gate count for small-scale Sycamore and heavy-hex backends.}

\vspace{-0.1in}

\begin{figure}[tb]
\begin{subfigure}[b]{0.23\textwidth}
    \centering
    \includegraphics[width=\textwidth]{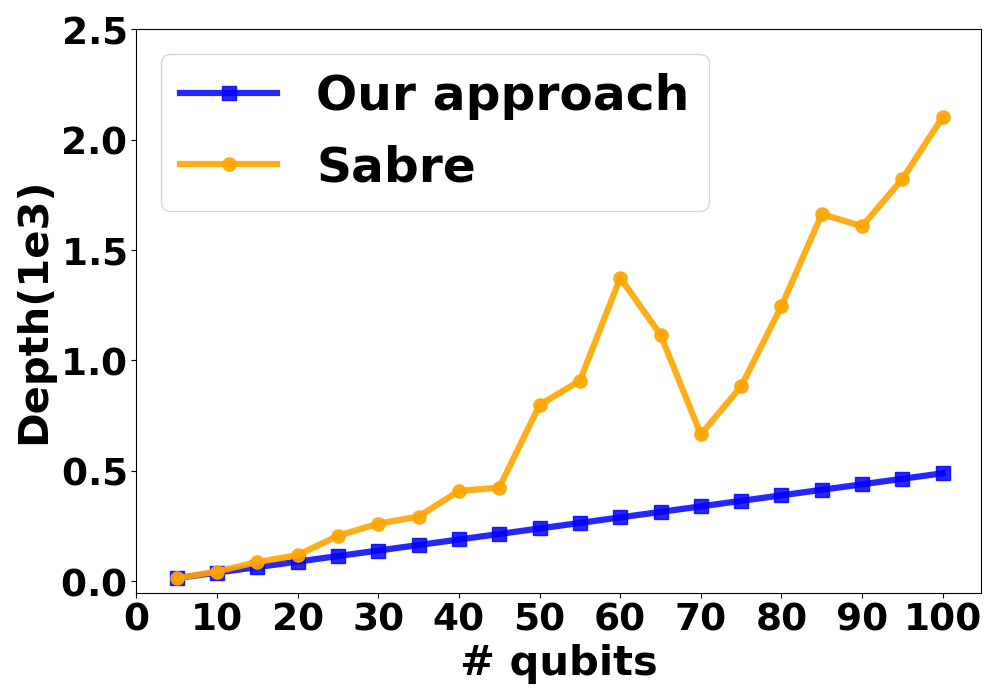}
    \vspace{-0.15in}
    \begin{small}
    \caption{Depth for Heavy-hex.}
    \end{small}
\end{subfigure}
\begin{subfigure}[b]{0.24\textwidth}
    \includegraphics[width=\textwidth]{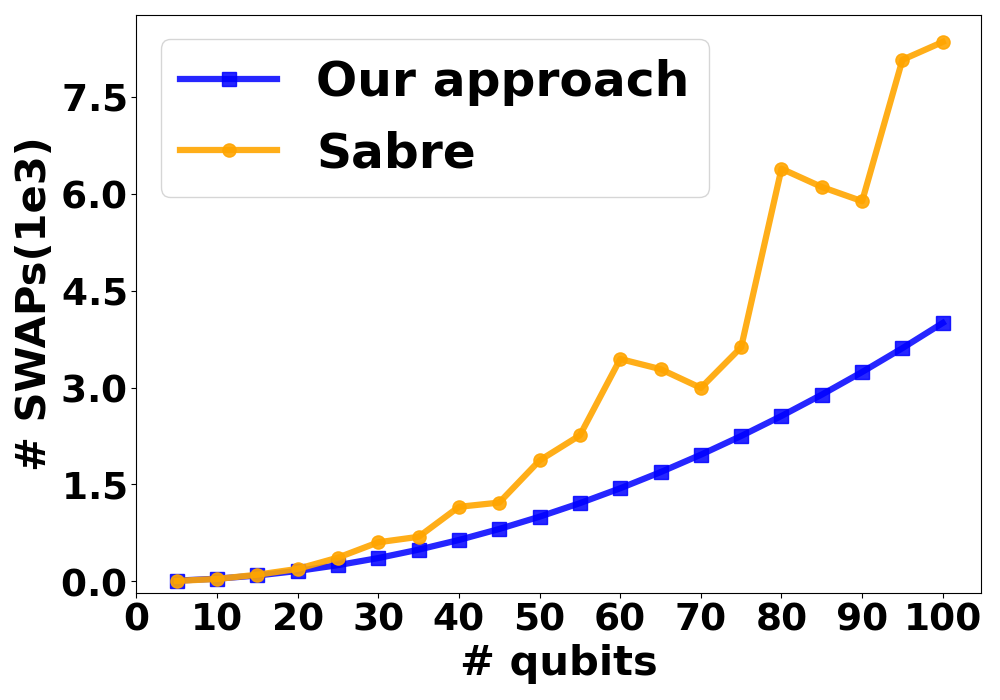}
    \vspace{-0.15in}
    \begin{small}
    \caption{\# SWAPs for Heavy-hex.}
    \end{small}
\end{subfigure}
\caption{Our approach vs. SABRE for Heavy-hex.}
\label{fig:heavy_hex}
\vspace{-0.15in}
\end{figure}

\begin{figure}[!t]
\begin{subfigure}[b]{0.23\textwidth}
    \centering
    \includegraphics[width=\textwidth]{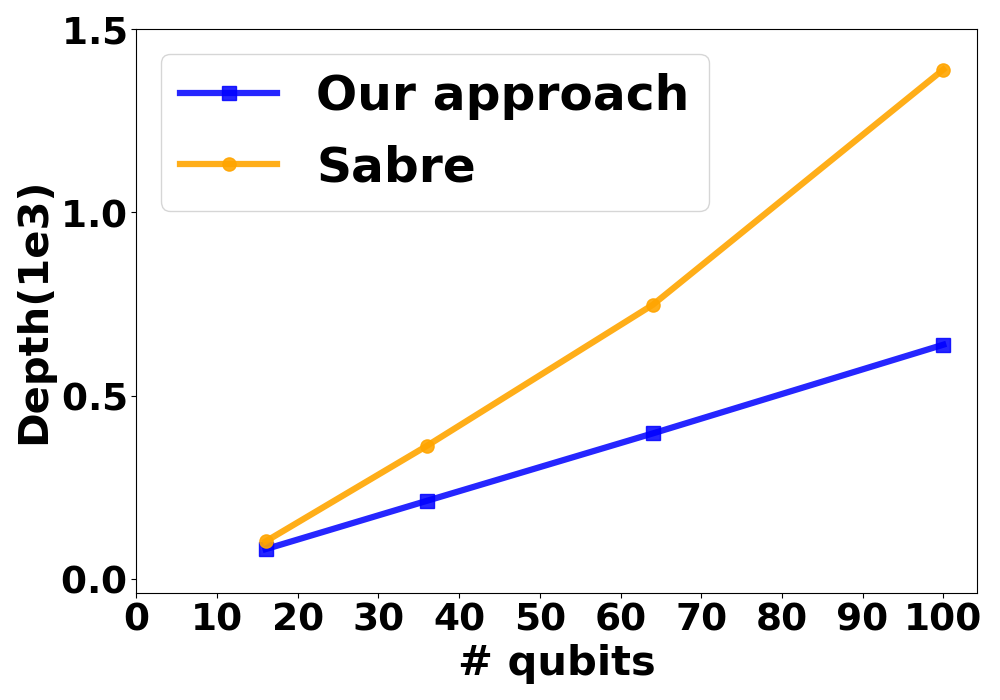}
    \vspace{-0.15in}
    \begin{small}
    \caption{Depth for Sycamore.
    }
    \end{small}
\end{subfigure}
\begin{subfigure}[b]{0.24\textwidth}
    \includegraphics[width=\textwidth]{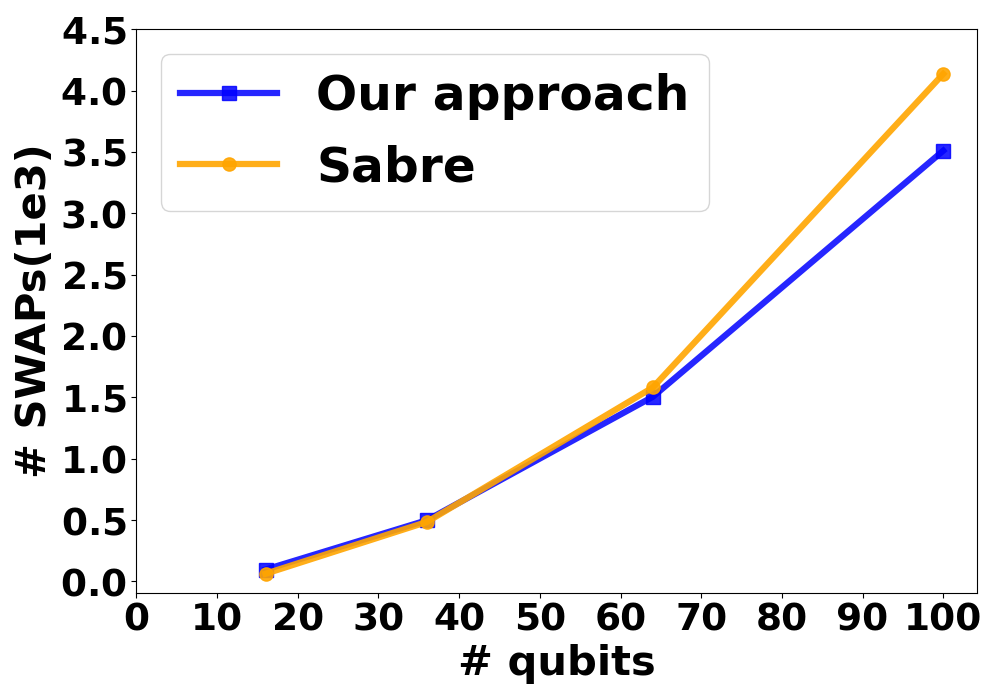}
    \vspace{-0.15in}
    \begin{small}
    \caption{\# SWAPs for Sycamore.
    }
    \end{small}
\end{subfigure}
\caption{Our approach vs. SABRE for Sycamore.}
\label{fig:Sycamore}
\vspace{-0.25in}
\end{figure}


\subsection{Generalizing Our Framework to Larger-scale QFT Kernel on FT Backends}
\vspace{-5pt}
FT backends are specifically designed for circuits with a large number of qubits, as only fault-tolerant systems have the capability to effectively manage the high error rates inherent in large-scale circuits. Conversely, NISQ systems are limited to managing error rates in smaller circuits. Therefore, whether our framework can be expanded to large-scale QFT relies on its performance on FT backends. In this part of the evaluation, we evaluate the performance of our approach, SABRE, and SATMAP on lattice surgery architecture. The size of $m$ is from 10 to 32, which leads to the number of qubits from 100 to 1024. We maintain the same aspects considered.

Due to the requirement of error correction, on the lattice surgery architecture, SWAP gate have different latencies on different links. 
We need to take into consideration these constraints. 
As SATMAP and SABRE lack the interface to configure such specific connections, we can't impose these constraints on them. 
Then we compare our approach against the version where all links are used for both baselines. 
Hence, if our approach can beat those baselines, we can conclude that our approach must be better than theirs. 
 
Due to the large qubit scale, SATMAP's evaluation always encountered a timeout across all QFT sizes, as shown in Table~\ref{table:expr}. 
SABRE generates the qubit mapping results faster than SATMAP at the cost of suboptimal results. 
The main reason is that SABRE incorporates several heuristics, some of which are greedy. The generated outcome may achieve even better performance (e.g., \# SWAP gate operations) than ours over small-scale architectures but as the size increases, its outcome becomes far worse than ours. 
Besides, the compilation time of SABRE increases proportional to the number of qubits in a particular architecture.

\begin{figure}[]
\begin{subfigure}[b]{0.23\textwidth}
    \centering
    \includegraphics[width=\textwidth]{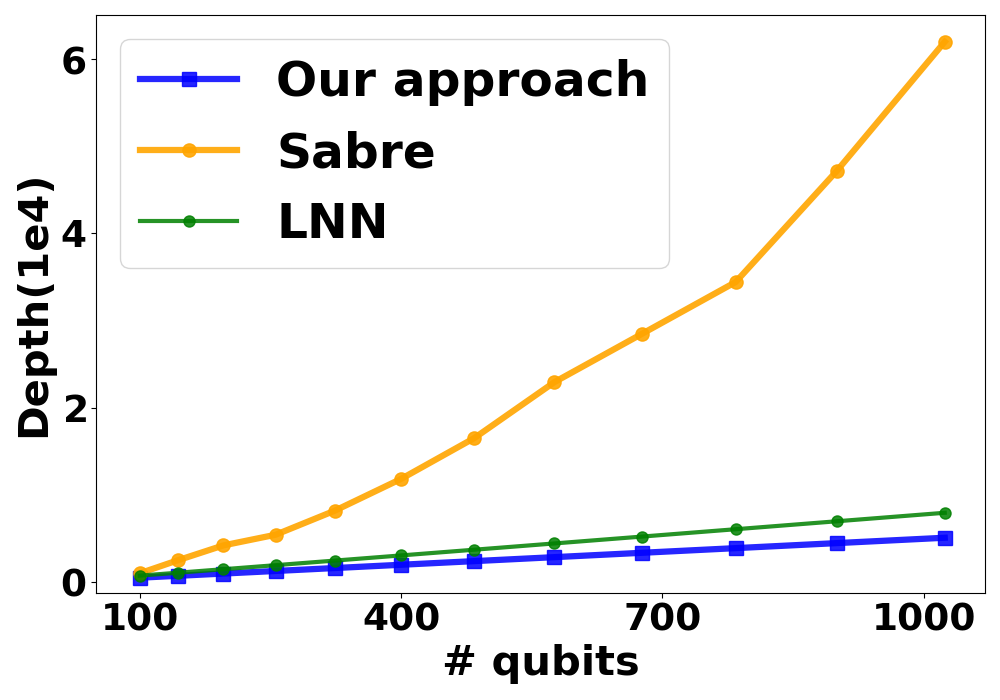}
    \vspace{-0.15in}
    \begin{small}
    \caption{Depth for Lattice Surgery.}
    \end{small}
\end{subfigure}
\begin{subfigure}[b]{0.24\textwidth}
    \includegraphics[width=\textwidth]{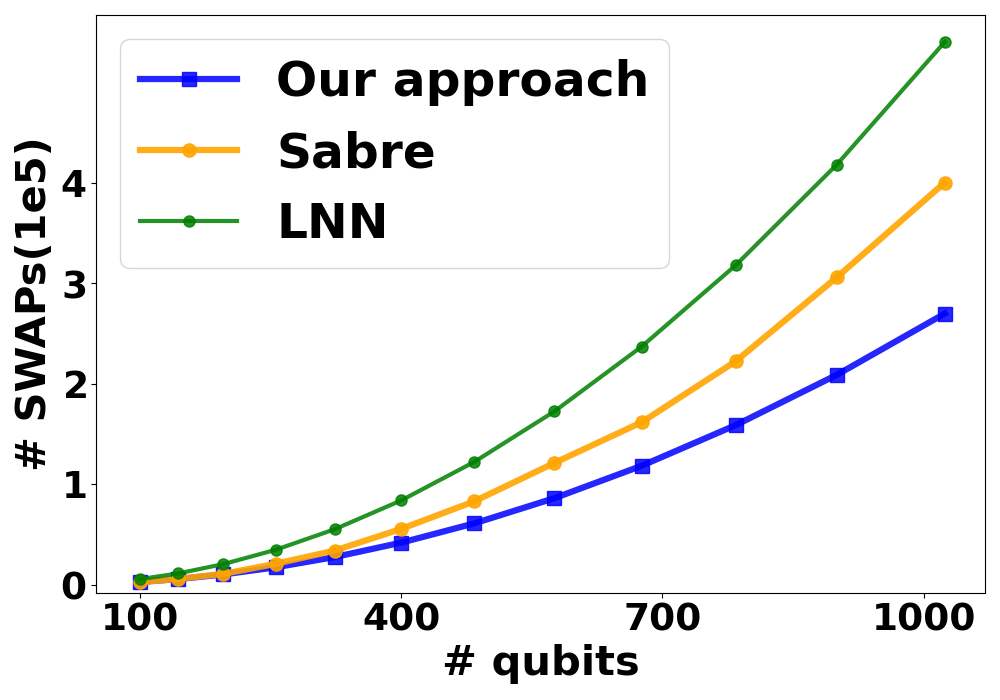}
    \vspace{-0.15in}
    \begin{small}
    \caption{\# SWAPs for Lattice Surgery. }
    \end{small}
\end{subfigure}
\caption{Our approach vs. SABRE for Lattice Surgery.}
\label{fig:lattice}
\vspace{-0.15in}
\end{figure}

To be precise, as for the depth (Fig. \ref{fig:lattice}), our approach is better than both SABRE and LNN among all sizes of qubits. The advantage becomes more significant as the size increases. 
Regarding the number of SWAP gates, our approach works better than SABRE when the number of qubits is larger than 144 and the advantage is also growing as the increasing of scale of QFT. Upon a detailed examination, for the scale of QFT comprising up to 1024 qubits, our approach demonstrates a significant advantage, achieving a depth cost that is roughly 92\% lower than SABRE's outcome. 


These results affirm that, despite the 
constraints favoring SABRE in the lattice surgery architecture (all links are used), our approach still delivers superior performance.

\textit{In conclusion, our framework exhibits enhanced performance in circuit depth and gate count for large-scale QFT circuits on FT backends. }

\vspace{-5pt}
\subsection{Scalability of the outcome} 
\vspace{-5pt}
Our approach relies on the systematic approach to complete the qubit mapping for QFT over diverse backends.
In contrast, this simplicity and scalability are not observed with methods like SABRE and SATMAP. 

SABRE, employs a look-ahead strategy to insert SWAPs, aiming to optimize not just for the immediate layer but future layers as well. 
It is difficult for us to find any common patterns that can be reused for a larger grid size.
The randomness of SABRE's output is in Fig.~\ref{fig:sabre} in Appendix. 
Getting output from SATMAP has already been challenging due to its long compilation time. Hence it is not a viable solution for scalable architectures.
\vspace{-5pt}

\section{Related Work}
\label{sec:relwork}
\vspace{-5pt}
Many studies focus on qubit mapping for a general class of applications \cite{siraichi+:cgo18, li+:asplos19, zhang+:asplos21, molavi+:micro22, tan+:iccad20, tan+:iccad22, niu_2020_tqe,zhang+:arxiv20, liu+:hpca2022swapsoptimization, liu+:qce2023qubitmapping}. A general-purpose compiler takes an arbitrary program and an arbitrary architecture as input, and produces a compiled circuit for this architecture. The issue with this approach is that every time the program size changes, 
for instance, the qubit number changes, 
the program needs to be recompiled. We focus on domain-specific qubit mapping and do not require the compiler to recompile the program when the input size changes. 
Our approach produces a linear-depth QFT circuit for both NISQ and FT backends.

There are also other domain-specific compilers for various applications including quantum approximate optimization algorithms (QAOA)\cite{alam+:dac20, alam+:micro20, lao+:isca22,jin_asplos23_qaoa},  variational quantum eigensolvers (VQE) \cite{li+:asplos22, li+:isca21, jin+:tetris}, and etc. 
For QFT, Maslov \etal \cite{maslov+:physreva07} for the first time shows a linear time solution on the linear nearest neighbor (LNN) architecture.  However, it is difficult to find a Hamiltonian path that connects all nodes in modern quantum architectures, limiting the applicability of this approach. Zhang \etal \cite{zhang+:asplos21} improved upon Maslov's \etal \cite{maslov+:physreva07} by discovering a linear-depth solution for a 2D grid with only two rows. However, 2xN grid architecture does not exist in modern architectures. Gao \etal \cite{gao+:arxiv2024_qft_ibm} proposed a similar approach to do qubit mapping for QFT over the IBM Heavy-hex NISQ devices. 

Leveraging program synthesis tools \cite{sketch} to do the compiler design for domain-specific applications~\cite{rmt} ~\cite{domino} ~\cite{cat} ~\cite{chipmunk} has already existed. 
However, to the best of our knowledge, our work is the first that demonstrates the usefulness of program synthesis for the compiler design domain of quantum computing, for the QFT kernel circuits.


\section{Conclusion}
\label{sec:conclusion}
\vspace{-5pt}
We propose a new QFT compilation framework for quantum application kernel over diverse quantum backends. Our approach outperforms the state-of-the-art approaches with less circuit depth and fewer gate count usage.

\section{Acknowledgements}
We extend our gratitude to the anonymous reviewers for their constructive and insightful feedback. This work was supported by grants from the Rutgers Research Council and NSF-FET-2129872. The opinions, findings, conclusions, and recommendations expressed in this material are those of the authors and do not necessarily reflect the views of our sponsors.






%
\bibliographystyle{plain}
\bibliography{allrefs}
\clearpage
\appendix
\renewcommand{\thesection}{\Alph{section}}
\subsection{Coupling graph generation for heavy-hex}
\label{appendix:coupling_graph_gen}
\begin{figure}[htb]
    \centering
    \includegraphics[width=0.49\textwidth]{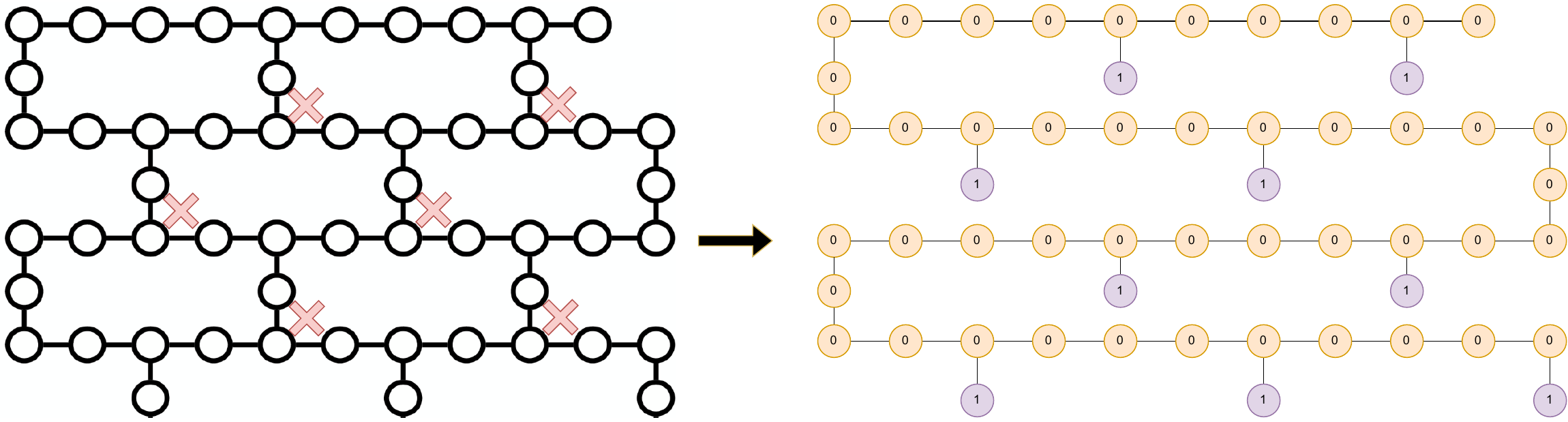}
    \caption{Turn heavy-hex (left-hand side) to our transformed coupling graph (right-hand side) with repeated patterns by removing some connection links (highlight by \textcolor{red}{X}). The transformed coupling graph consists of a major line (labeled by 0) with dangling points (labeled by 1).}
    \label{fig:IBM}
\end{figure}

As a concrete example shown in Figure~\ref{fig:IBM}, we delete some connection links from heavy-hex architecture to form the simplified coupling graph with one main line and dangling points. 

\subsection{Heavy-hex Special Case}
\label{appendix:heavyhex_time}
In special case, we represent heavy-hex as a coupling graph. In this graph, there is one dangling point for every 4 points in the main line. 

\begin{figure*}[]
    \centering
    \includegraphics[width=0.8\textwidth]{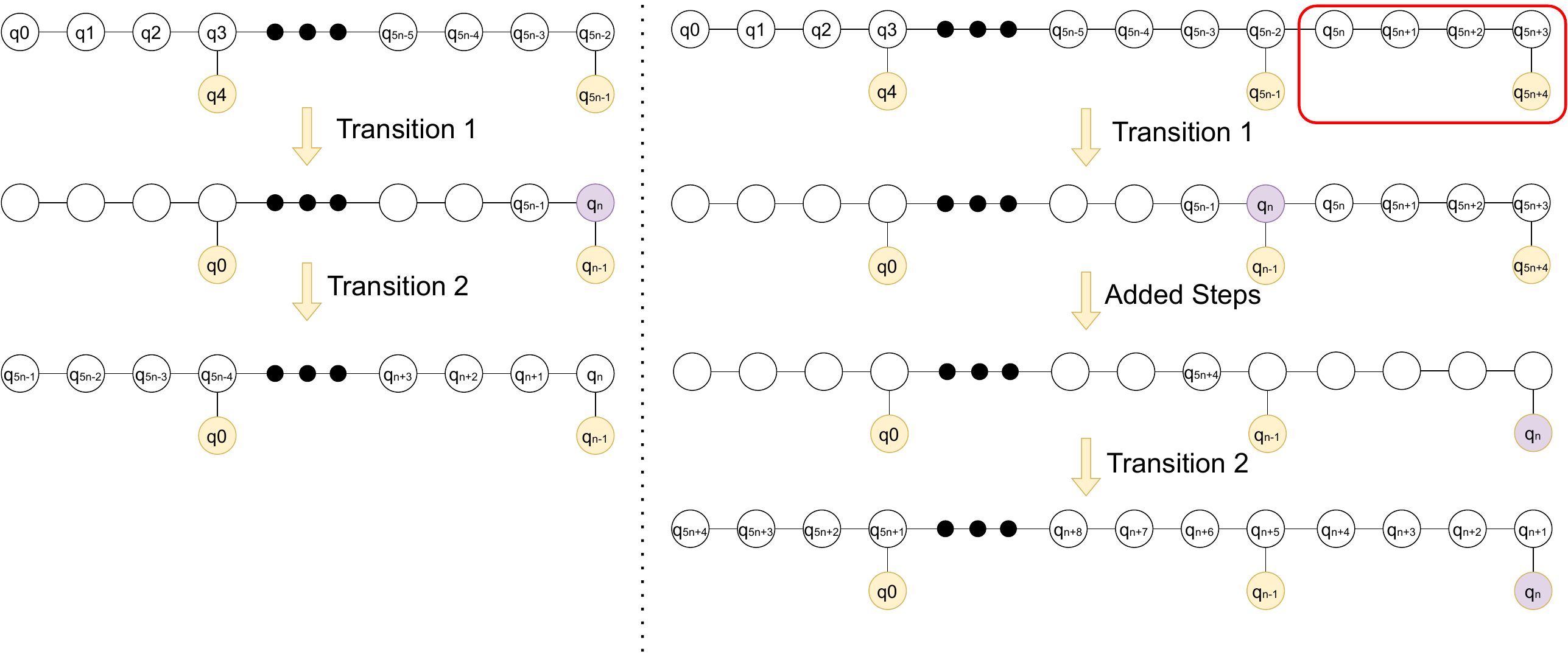}
    \vspace{-0.3cm}
    \caption{The extra steps for one newly added group. The LHS is the QFT mapping for architecture with n groups while the RHS has (n+1) groups. They share the same cycles in transition 1 and transition 2 but the RHS has to use added steps to complete additional gate operations for newly added qubits.}
    \label{fig:extra_steps_for_new_group}
\end{figure*}

Figure~\ref{fig:extra_steps_for_new_group} decouples the QFT mapping of this special-case heavy-hex architecture into multiple states and captures the extra steps for a newly added group in the architecture. 
To obtain extra steps, we should pay attention to the operation over the \textbf{physical qubits} rather than the logical qubits. 
The LHS of Figure~\ref{fig:extra_steps_for_new_group} divides the QFT mapping for 5n qubits into three states: initial state, the state when the qubit with the biggest number $q_{5n-1}$ is one step before $q_{n}$, and the final state. 
Let's use $Q_{k}$ to label the physical qubit where $q_{5n-1}$ sits after transition 1.
Compared with the version where there are 5n + 5 qubits in the RHS, the added steps occur to move $q_{n}$ to its final destination after transition 1 and move the qubit with the biggest number $q_{5n+4}$ to physical qubit $Q_{k}$. The remaining steps overlap.

\begin{figure}[]
    \centering
    \includegraphics[width=0.49\textwidth]{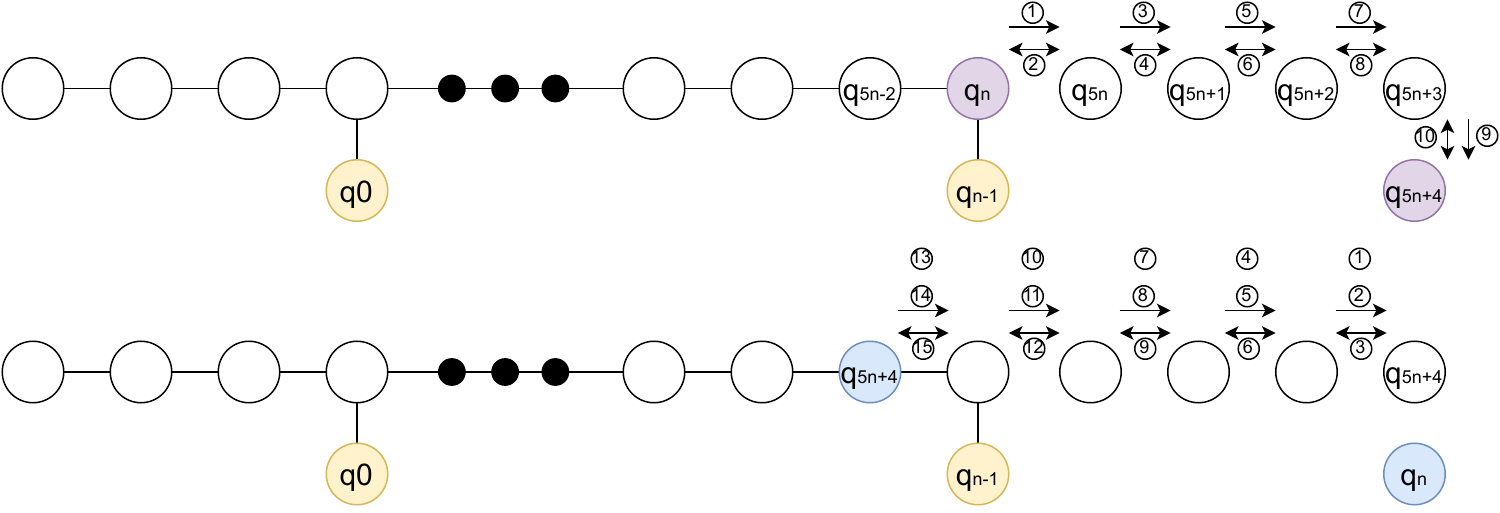}
    \vspace{-0.1cm}
    \caption{Extra steps required for one added qubit group. The top figure shows the steps in the first phase, while the bottom figure shows that in the second phase.}
    \label{fig:extra_steps}
\end{figure}

To calculate the concrete number of added steps, we visualize them in 2 phases in Figure~\ref{fig:extra_steps}. 
The first phase is to move $q_{n}$ forward until its final destination is located at the end of dangling positions; 
the second phase is to move $q_{5n+4}$ backward to the places where $q_{5n-2}$ sits before phase 1. 
The first phase takes 10 steps to move $q_{n}$ forward until the last location of the dangling positions, and another 15 steps are necessary in the second phase to move $q_{5n+4}$ backward to where $q_{5n-2}$ was before. An extra 25 steps are required to complete the QFT mapping for one added group with 5 qubits. Therefore, as the number of qubits grows, the time complexity would be 5N $+O(1)$, where N is the total number of qubits in the heavy-hex architecture.

\subsection{Heavy-hex General Case}
\label{appendix:heavyhex_time_gen}
Different from the special case, the general case heavy-hex \textit{does not} require one dangling point for every 4 points in the main line. 
To calculate the time complexity, we decompose the whole QFT mapping process into 2 main parts and do the calculation for each part individually. 

Part I: QFT mapping in the major line only. 
Zhang \etal \cite{zhang+:asplos21} show that QFT mapping over an LNN consists of phases with only CPHASE gate operations and with SWAP operations. 
In a straight line with N1 qubits, there are 2*N1-3 CPHASE cycles and 2*N1-3 SWAP cycles. Therefore, the total number of cycles is 4*N1-6.

Part II: interaction between the major line and the dangling points. 
This part can be divided into 2 categories. 
The first category is related to the steps required to SWAP the qubit in the major line with the dangling point above. Specifically, since there are N2 dangling points, the final mapping shows that qubits from $q_0$ to $q_{N2-1}$ should finally be located in the dangling position. This means a SWAP operation is required to change the location between $q_{i}$ and the dangling qubit above when it reaches the ``T junction" above its final destination from the left. 
In addition to the SWAP operation, there is a CPHASE operation before, so for each of these N2 qubits, it needs an extra 2 phases to do such interaction, and the total time complexity is 2*N2;
the second category is related to the CPHASE gate operation between dangling points and major lines where the dangling qubit is used as the control qubit. These steps should be added when the major line is doing LNN QFT mapping. As we can see from the existing QFT mapping over LNN, there will be interleaving between CPHASE and SWAP cycles. The CPHASE cycle will do CPHASE gate operations, not changing the relative qubit position while SWAP cycle will change some qubits' position. Therefore, the extreme case is that each time we have a SWAP cycle, we need to stop for one cycle to do the CPHASE gate operations in the second category. The total time complexity is 2*N1-3.

By summing up the complexity from 2 parts, we can get the upper bound, which is 6N1 - 9 + 2N2. Since there is a known fact that $N1 + N2 = N$, $N1 \leq N$, and $N2 \leq N$, the upper bound time complexity is 2N+4N1+O(1) $\leq$ 6N+O(1).

\subsection{Final Qubit Mapping for Heavy-hex}
\begin{figure}[htb]
    \centering
    \includegraphics[width=0.45\textwidth]{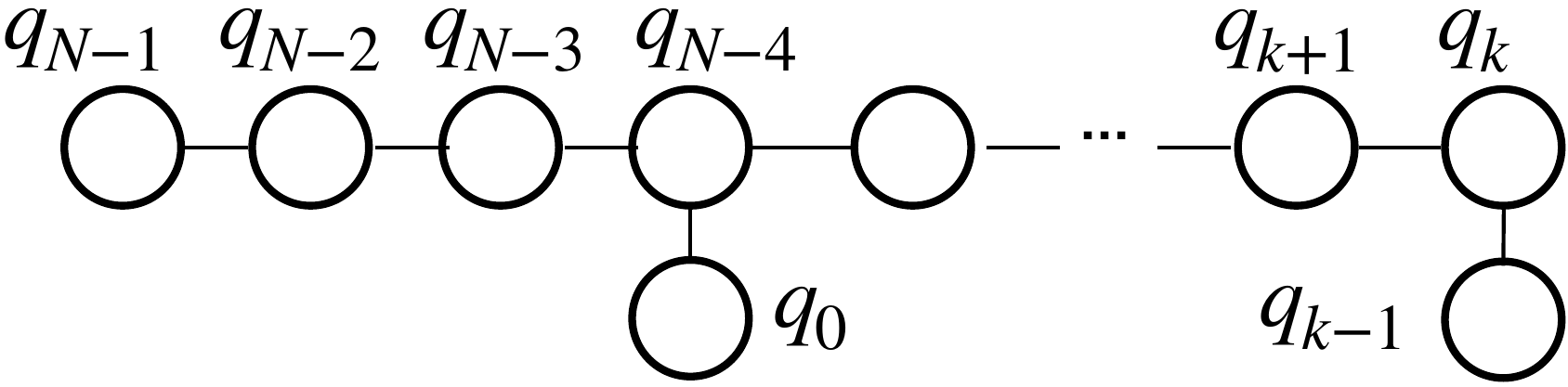}
    \caption{Final mapping for Heavy-hex with N qubits.}
    \label{fig:final_mapping_heavy_hex}
\end{figure}

The final qubit mapping for heavy-hex is shown in Fig.~\ref{fig:final_mapping_heavy_hex}.

\subsection{Program synthesis for Inter-Unit of the Google Sycamore Architecture}
\label{appendix:program_synthesis}
\begin{figure}[htb]
    \centering
    \includegraphics[width=0.3\textwidth]{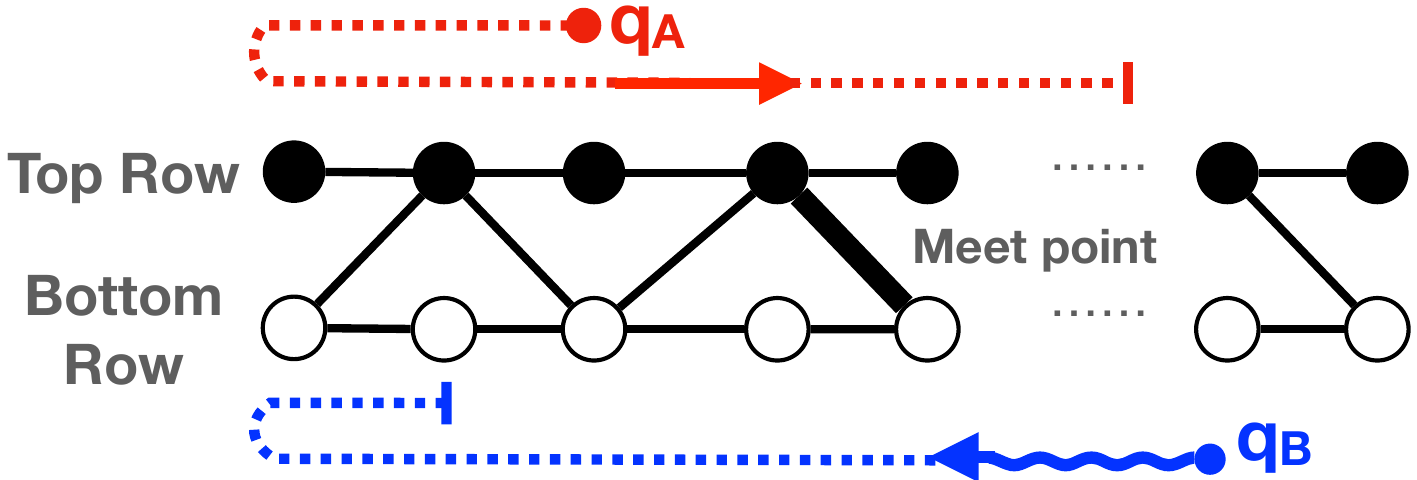}
    \caption{Condition for 2 qubits in two different units to meet in Google Sycamore. }
    \label{fig:sycamorepathcross}
\end{figure}

In the two-unit case, it is as if there are two lines. We refer to them as the top row and the bottom row as shown in Fig. \ref{fig:sycamorepathcross}. If a qubit in the top row moves along a path, and a qubit in the bottom row moves along a path, and somehow they meet through a link between these two rows, then we can run a CPHASE gate. The uniqueness in Sycamore's inter-unit case is that there is a connection between one qubit on the top row and the qubit on the bottom row \emph{if the two qubits' column index differ by 1}, as shown in Fig. 
\ref{fig:sycamorepathcross}.  

The challenge is how to move qubits in the top row and bottom row, and let them meet. Our guess is that we may sync the movement of qubits on the top row and bottom row. The reason is that if qubits are on a single line, any two qubits will have a distance of 1 at some time point, if they follow the movement pattern in LNN. This is analog to the case that the column index differs by 1 for two linked qubits from the top row and bottom row respectively.


But one also has to be careful, in the Sycamore architecture, two vertices having the same column index on the top and bottom row, do not have a direct connection between them. Moreover, between two units, there are $row\_size - 1$ links.  Fortunately, it is simple to figure out a solution for CPHASE gates between pairs of qubits in the same column. We can SWAP one of them horizontally with its neighbor at the top (bottom) row, keep the other one on the bottom (top) row unchanged, run a CPHASE gate, and then revert the qubit back to its original location with the same SWAP.

Hence, we really want to ask one question: \textit{if we exclude all CPHASE gate operations for qubit pairs in the same column initially, could program synthesis solve the rest gate operations assuming we sync the steps of the top and bottom row?} 
If there is such a solution that accords with our assumption, it might be possible for us to generalize something useful from it.


\textbf{\emph{Specification}}  First of all, we enable the program synthesis tool to exclude CPHASE gate operations between qubits that are in the same vertical lines initially (e.g., in Fig.~\ref{fig:googlesycamore_unit} all ($q_i$, $q_{i + 6}$) pairs). The excluded gate operations will be fixed using the way mentioned above.

\begin{figure*}[htb]
    \centering
    \includegraphics[width=\textwidth]{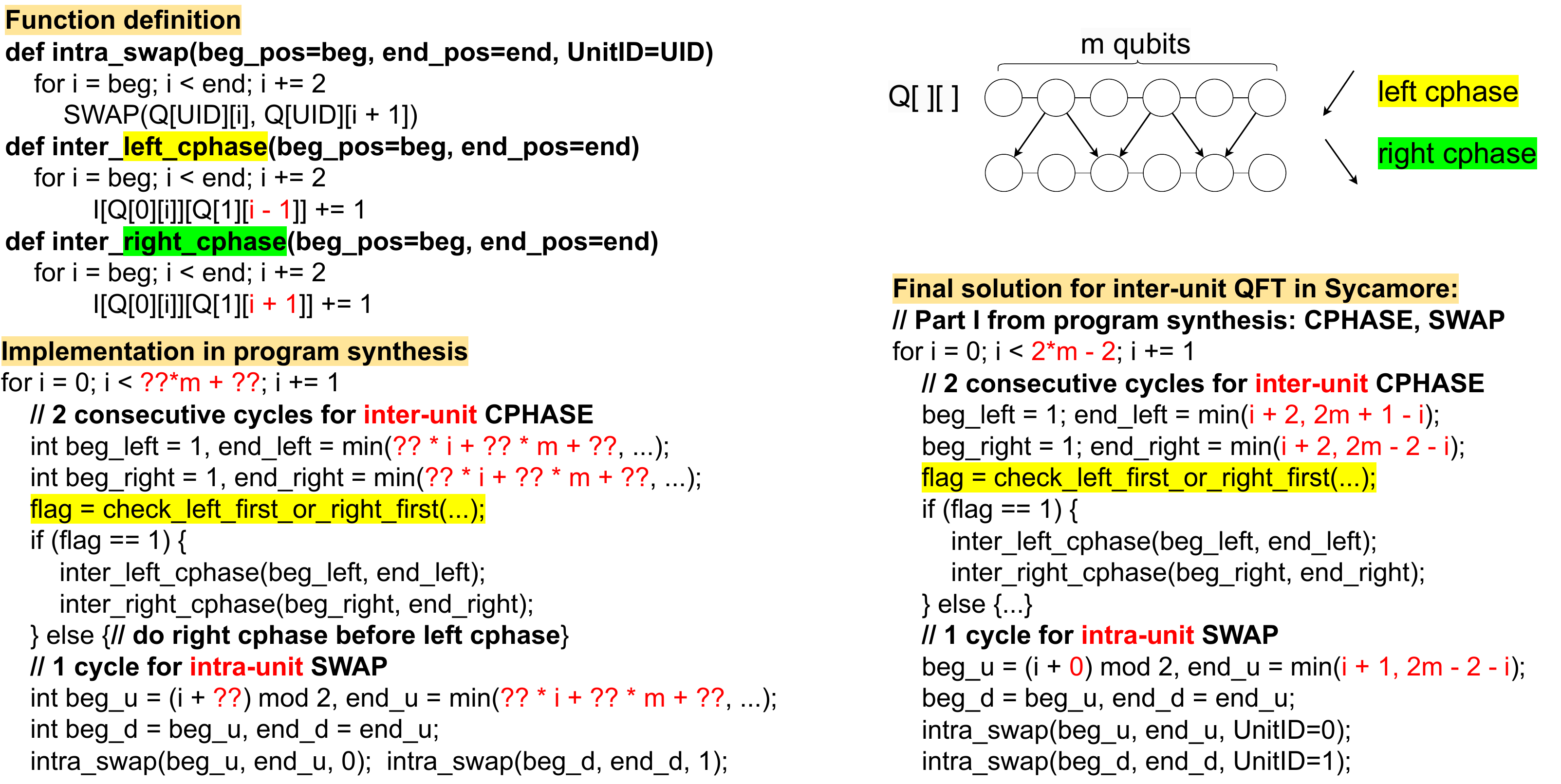}
    \caption{Program synthesis implementation and outcome for Sycamore using SKETCH.}
    \label{fig:impl_out}
\end{figure*}

Additionally, we allow a CPHASE gate to happen more than once between two qubits just for synthesis purposes. This is to account for the potential boundary case in a loop, where there might be a head and tail condition that is different from the repeated steps in the middle. The head and tail conditions are boundary conditions.  Later, if the program synthesis engine successfully figures out the repeated steps, we can fix the solution by manually setting the head and tail components of the loop, and ensuring only one CPHASE between each pair of qubits.  



\textbf{\emph{Implementation}}  The implementation is shown in Fig. \ref{fig:impl_out}. The intuition is multi-fold. First, we maximize the utilization of all links between 2 units. We run all CPHASE gates between two units as long as their dependence is resolved. Since the links between two units form a line, it needs at least two steps to run all of them (or a consecutive subset of them), we divide them into the left and right CPHASE gates.

Second, the starting point for the SWAPs in one loop iteration should be different from that in the next loop iteration. Otherwise, it means nothing but to undo the previous SWAP; therefore, we use the ``mod" function to reflect this feature. 
\begin{figure*}[]
    \centering
    \includegraphics[width=0.95\textwidth]{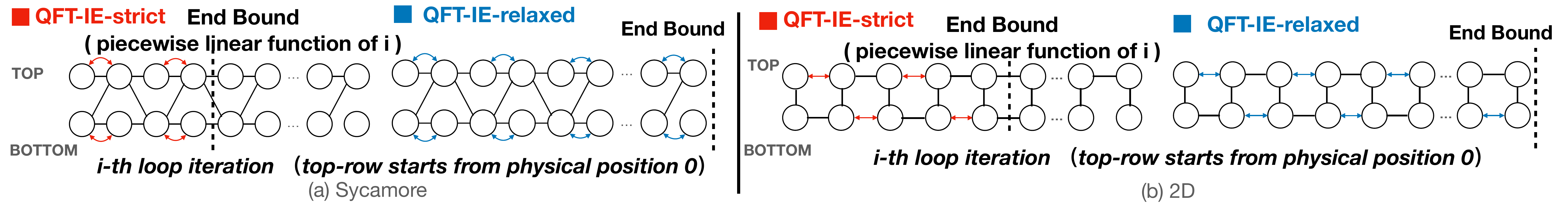}
    \caption{QFT-IE-strict and QFT-IE-relaxed in Sycamore and 2D grid. Note that the top row starts at position 0 in this example, but it in fact alternates between 0 and 1 in consecutive loop iterations for both Sycamore and 2D grid.  We omit the 1 case due to the space limit. (a) Sycamore, and (b) 2D grid.  }
    \label{fig:QFT-IE-Sycamore-strict-relaxed}
\end{figure*}

Third, inspired by the SWAP bound in the LNN case in Fig. \ref{fig:LNN5}, where a triangle shape covers the set of valid SWAP steps.  We ``guess" that the bounding point for the SWAPs in one layer might follow a piecewise linear function of loop induction variable represented using the ``min(...)" function. The \emph{min(...)} function may take more than 2 arguments, but we find two are enough in our case. 

\textbf{QFT-IE-strict and QFT-IE-relaxed Solutions}
The inter-unit QFT-IE-strict solution is presented in Fig. \ref{fig:impl_out} in pseudocode. The repeated steps are visualized in Fig. \ref{fig:QFT-IE-Sycamore-strict-relaxed}. 

The inter-unit QFT-IE-relax solution is below:
\begin{footnotesize}
\begin{verbatim}
    for (i = 0; i <= m; i += 1)
       CPHASE on all inter-unit connections   
      // Intra-unit swap
      beg = i mod 2
      intra_swap(beg_pos=beg, end_pos=m, UnitID=0)
      intra_swap(beg_pos=beg, end_pos=m, UnitID=1)
\end{verbatim}
\end{footnotesize}

The repeated steps are shown in Fig. \ref{fig:QFT-IE-Sycamore-strict-relaxed} (a). In the implementation of the piecewise linear functions, the QFT-IE-relaxed version in fact chooses to use the same function for two parameters in min(). But QFT-IE-strict chooses two different linear functions. The QFT-IE-relaxed version is two times faster than the QFT-IE-strict version.

One extra benefit of our solution is that both QFT-IE-strict and QFT-IE-relax will mirror the position of all qubits within a unit. Hence it facilitates further processing that for QFT-IA. In Google Sycamore, since QFT-IA uses the LNN solution, the qubits must be placed in natural number ordering in the LNN, or the reversed natural number ordering.


\subsection{Output analysis for SABRE and SATMAP}
\begin{figure}[htb]
    \centering
    \includegraphics[width=0.46\textwidth]{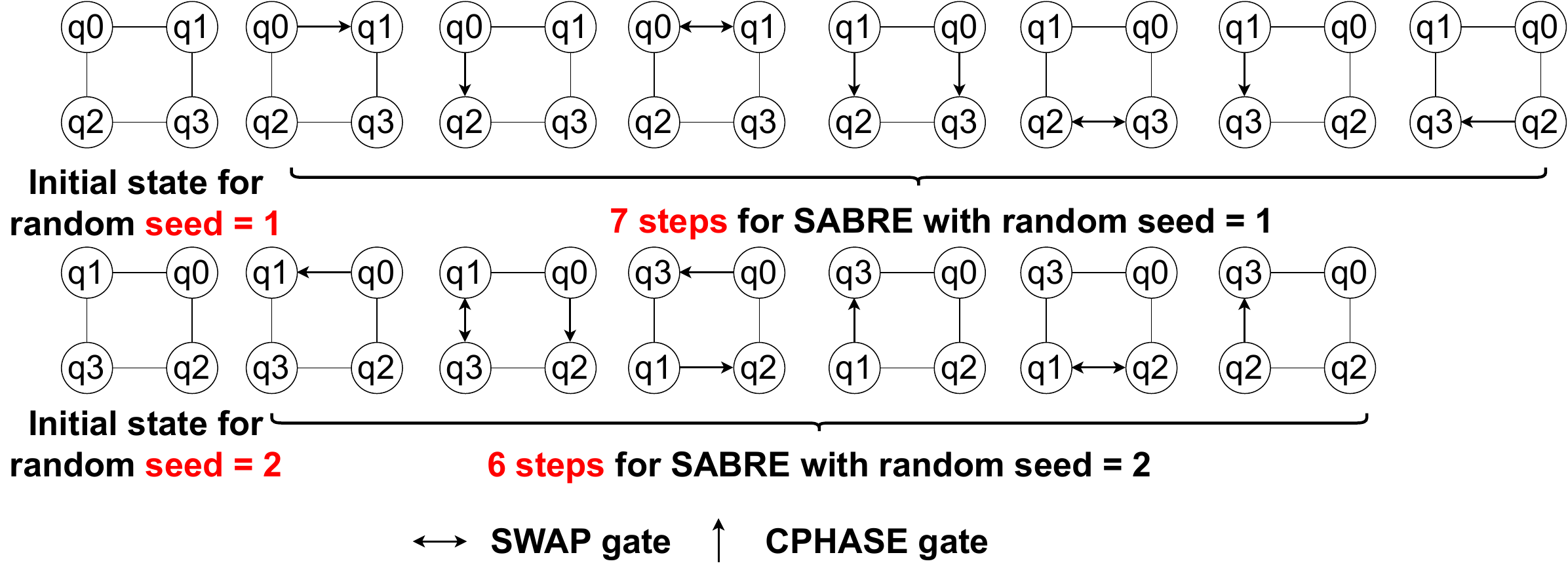}
    \caption{Randomness in SABRE's output.}
    \label{fig:sabre}
\end{figure}

The outcomes produced by SABRE are influenced by a random seed, which leads to varying outcomes for each compilation.
For instance, Fig.~\ref{fig:sabre} presents the output of SABRE using different random seeds for QFT over a 2 by 2 grid size. The outcome might vary from seed to seed in terms of initial mapping, gate operation order, and even the final depth to complete QFT.

\subsection{Program Synthesis for Inter-unit of the Regular 2D NxN Grid }

The intuition is contrary to that of the inter-unit in Google Sycamore. Recall that we ``guessed" in Google Sycamore, that the top row and bottom should move in sync, but now this should not work.   

\begin{figure}[htb]
    \centering
    \includegraphics[width=0.45\textwidth]{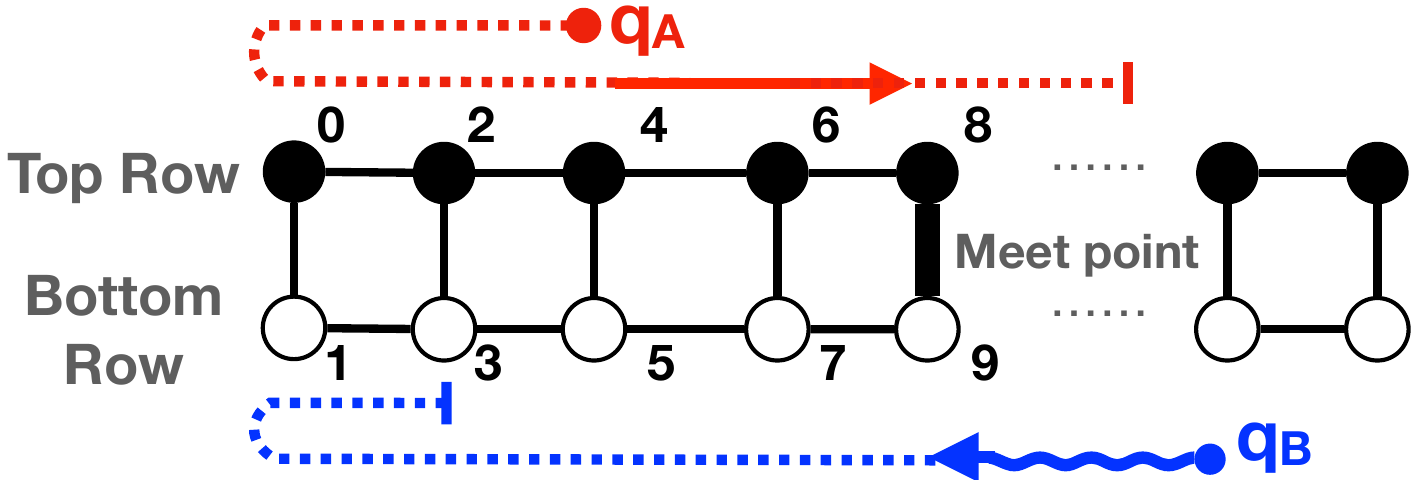}
    \caption{Trajectory of two points from the top row and the bottom row.}
    \label{fig:pathcross}
\end{figure}

If we look at path crossing in Fig. \ref{fig:pathcross}, now the two qubits on the top row and bottom row should meet at exactly the same column, as opposed to two columns of distance 1. This is because a 2D grid has direct vertical links between qubits in the top row and the bottom row. 

In this case, the top-row qubits and bottom-row qubits should not move in the same way. Otherwise, for any column at any time instant, the top qubit's neighbor in the bottom row is always the same. This makes it impossible to do all-to-all interaction between these two rows.

Now our guess is that each qubit should still travel along a path, but there must be a difference between the starting points of movement (SWAPs) in the top row and the bottom row. Two qubits on different rows will have their paths cross at some point (meeting in the same column).  This leads to our program synthesis model as follows. 

\textbf{Specification}
The specification is similar to that of Google Sycamore. The only difference is that we do not exclude the qubit pairs in the vertical links at initial qubit mapping.

\begin{figure}[htb]
    \centering
\includegraphics[width=0.48\textwidth]{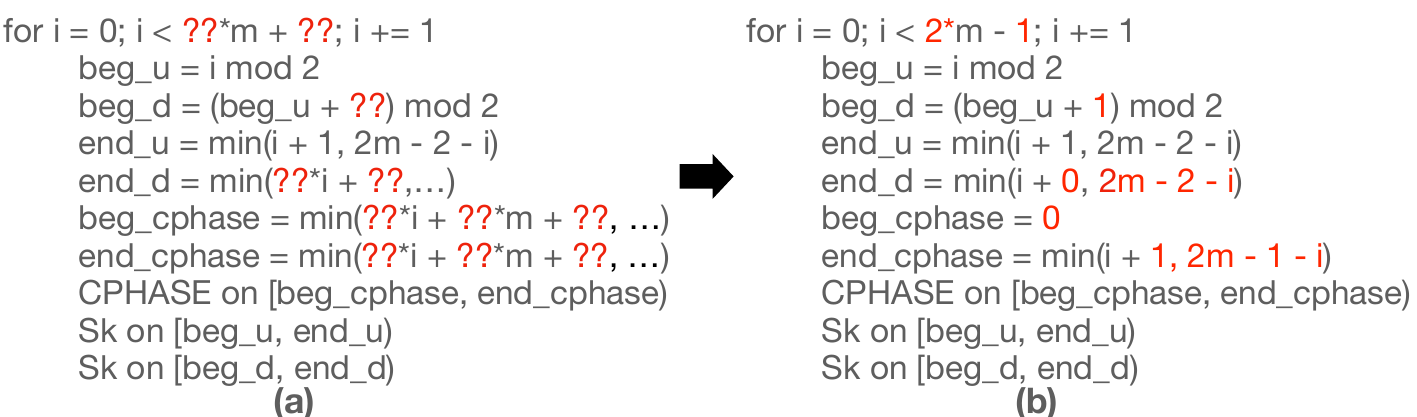}
    \caption{Synthesized code for inter-unit of the 2D regular grid in the strict version. }
    \label{fig:syn2dIE}
\end{figure}

\textbf{Implementation.} With the above conditions, our formulated code shape is in Fig. \ref{fig:syn2dIE}.  

In the code in Fig. \ref{fig:syn2dIE}, ``beg\_u" and ``end\_u" respectively represent the begin and end location for the SWAPs in the top row. ``beg\_d" and ``end\_d" respectively represent the begin and end locations for the SWAPs in the bottom row. CPHASE gates operate on a range of vertical links, and the range is a linear function of the loop induction and loop invariant variables too. $m$ is the unit size.

We specifically add a difference between the starting location of the top-row SWAP, and that of the bottom-row SWAP, using the modular function ``beg\_u + ??" mod 2.  We simplified the loop by fixing the movement of the top-row as that in QFT LNN. In this case, the simple code shape actually worked. It found a solution successfully, shown in Fig. \ref{fig:syn2dIE} (b). 

\begin{figure}[htb]
    \centering
    \includegraphics[width=0.47\textwidth]{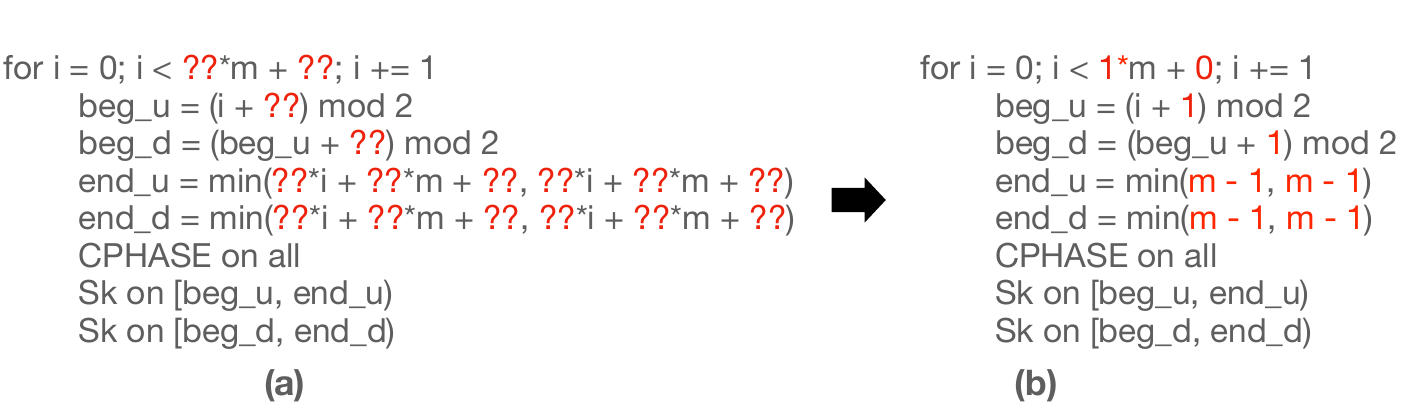}
    \vspace{-5pt}
    \caption{2-unit interaction for the relaxed order QFT-IE.}
    \label{fig:syn2drelaxed}
    \vspace{-0.1in}
\end{figure}

We took a step further and chose not to fix the movement in the top row. Rather, we use piecewise linear functions to specify the beginning and ending locations for intra-row SWAP, as shown in Fig. \ref{fig:syn2drelaxed}. Using this approach, we found a solution for QFT-IE-relaxed as shown in Fig. \ref{fig:syn2drelaxed} (b). The repeated step is visualized in Fig. \ref{fig:QFT-IE-Sycamore-strict-relaxed} (b).

\end{document}